\documentclass[sigconf]{acmart}
\AtBeginDocument{%
  }

\usepackage{amsmath,amsfonts}
\usepackage{amssymb}
\usepackage{algorithm}
\usepackage{algorithmicx}
\usepackage[noend]{algpseudocode}
\usepackage{graphicx}
\usepackage{textcomp}
\usepackage{enumerate}
\usepackage{multirow}
\usepackage{makecell}
\usepackage{xspace}
\usepackage{subfig}
\usepackage{booktabs} 
\usepackage{fancyhdr}
\usepackage{amsthm}
\usepackage{bm}
\usepackage{mathrsfs}
\usepackage{utfsym}
\usepackage{url}
\usepackage{diagbox}
\usepackage{float}

\usepackage{anyfontsize}

\newtheorem{definition}{Definition}

\newcommand{\sysname}{\textsc{CrossMPI}\xspace}

\newcommand{\et}{\textit{et al.}}

\newcommand{\rnum}[1]{\uppercase\expandafter{\romannumeral#1}}

\usepackage{tabulary}
\usepackage{colortbl}
\definecolor{darkred}{RGB}{255,191,191}
\definecolor{darkgreen}{RGB}{182,228,182}

\newcommand{\mypara}[1]{\noindent{\bf {#1}.} \xspace}

\usepackage{enumitem}
\setlist[itemize]{leftmargin=*}

\usepackage{pifont}

\newcommand{\refappendix}[1]{\hyperref[#1]{Appendix~\ref*{#1}}}

\begin{document}

\title{A Cross-Modal Prompt Injection Attack against Large Vision-Language Models with Image-Only Perturbation}

\author{Hao Yang}
\affiliation{%
  \institution{Xidian University}
  \city{Xi'an}
  \state{Shaanxi}
  \country{China}
}
\email{yangh@stu.xidian.edu.cn}
\orcid{0009-0002-1274-4325}

\author{Zhuo Ma}
\authornote{Corresponding author.}
\affiliation{%
  \institution{Xidian University}
  \city{Xi'an}
  \state{Shaanxi}
  \country{China}
}
\email{mazhuo@mail.xidian.edu.cn}
\orcid{0000-0001-6023-2864}

\author{Yang Liu}
\affiliation{%
  \institution{Xidian University}
  \city{Xi'an}
  \state{Shaanxi}
  \country{China}
}
\email{bcds2018@foxmail.com}
\orcid{0000-0002-6276-1468}

\author{Yilong Yang}
\affiliation{%
  \institution{Xidian University}
  \city{Xi'an}
  \state{Shaanxi}
  \country{China}
}
\email{yilongyang@xidian.edu.cn}
\orcid{0000-0002-2811-2667}

\author{Guancheng Wang}
\affiliation{%
  \institution{Xidian University}
  \city{Xi'an}
  \state{Shaanxi}
  \country{China}
}
\email{wangguancheng@stu.xidian.edu.cn}
\orcid{0009-0005-4070-0153}

\author{JianFeng Ma}
\affiliation{%
  \institution{Xidian University}
  \city{Xi'an}
  \state{Shaanxi}
  \country{China}
}
\email{jfma@mail.xidian.edu.cn}
\orcid{0000-0003-4251-1143}




\begin{abstract}
Large vision-language models (LVLMs) have emerged as a powerful paradigm for multimodal intelligence, but their growing deployment also expands the attack surface of prompt injection. 
Despite this growing concern, existing attacks still suffer from a critical limitation: the injected prompt for one modality only steers the model's interpretation of that singular input.
Alternatively, these attacks remain \emph{multimodal} but fail to achieve \emph{cross-modal} prompt perturbation.
To bridge this gap, we introduce a novel \underline{cross}-\underline{m}odal \underline{p}rompt \underline{i}njection attack \sysname, which can steer the model's interpretation of both textual and visual inputs via image-only prompt injection.
Our design is underpinned by the following key breakthroughs.
First, we turn the focus of the injected prompt perturbation optimization from the visual embedding space (typically with only $10^5$ parameters) to the model hidden state space (for multimodal information integration and with $10^7$ parameters).
Then, two strategies are adopted to mitigate the optimization challenges posed by the larger parameter space.
To constrain the optimized model parameter space, we introduce a layer selection strategy that identifies the layers most critical to multimodal integration.
Interestingly, deviating from the past experience, our analysis reveals that the optimal layers for LVLM prompt perturbation reside in the middle of the model rather than the last.
To constrain the image perturbation space, we propose a new distance-decremental perturbation budget assignment strategy that allocates budgets decrementally as the pixel distance to semantic-critical regions increases.
Extensive experiments across multiple LVLMs and datasets show that our method significantly outperforms baseline approaches.

\end{abstract}



\renewcommand\footnotetextcopyrightpermission[1]{}
\settopmatter{printacmref=false} 



\maketitle

\section{Introduction}
\label{sec:Introduction}
\begin{figure}[ht]
    \centering 
    \includegraphics[width=\linewidth]{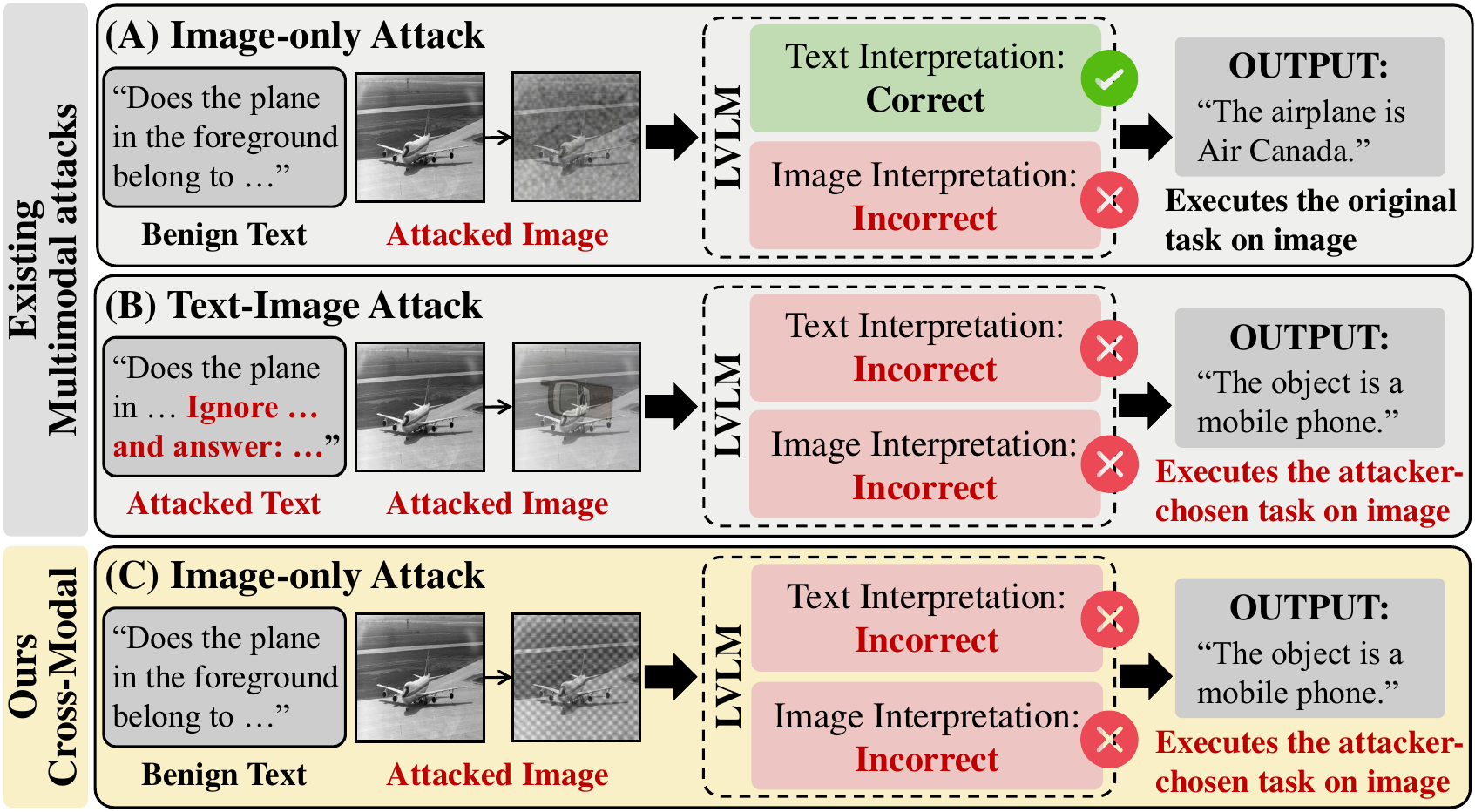}
    \caption{Comparison of prompt injection paradigms in LVLMs. (A) Image-only attacks perturb the image to change visual perception, while the model still follows the benign text task. (B) Text-image attacks change the task execution by perturbing both the text prompt and the image. (C) Our cross-modal prompt injection perturbs only the image but changes how the model interprets the unchanged text prompt.}
    \label{fig: intro}
    \Description[Comparison of LVLM prompt injection paradigms]{A three-panel comparison showing image-only perceptual attacks, multimodal attacks with explicit malicious instructions, and the proposed image-only cross-modal prompt injection attack.}
\end{figure}
Recently, Large Vision-Language Models (LVLMs)~\cite{achiam2023gpt, chen2024sharegpt4v, dai2023instructblip, liu2023visual, wang2024cogvlm, zhuminigpt, wu2024next, zhang2024mm} have rapidly advanced and emerged as a powerful paradigm for achieving unified multimodal intelligence.
Unlike Large Language Models (LLMs) that focus solely on the text modality, LVLMs jointly model and reason over both visual and text inputs, enabling a wide range of complex cross-modal tasks, such as visual question-answering~\cite{tsimpoukelli2021multimodal, li2023blip, alayrac2022flamingo}, text-to-image generation~\cite{nichol2022glide, ramesh2022hierarchical, rombach2022high}, and creative content creation~\cite{zhuminigpt}.
By effectively integrating the complementary information encoded across different modalities, these models are capable of producing reasoning and decision-making that are more semantically coherent and context-aware, and they demonstrate substantial potential for continued expansion in the pursuit of general artificial intelligence and real-world applications.

Alongside their growing capabilities, recent studies have shown that LVLMs inherit and amplify security vulnerabilities originally observed in LLMs, among which \textit{prompt injection attacks} pose a particularly severe threat.
Prompt injection~\cite{liu2024formalizing, shi2024optimization, pasquini2024neural, toyertensor} introduces malicious instructions that override or conflict with the user's original intent, causing the model to execute an attacker-chosen injected task rather than the intended target task.
In LLM-based systems, attackers typically achieve this by inserting carefully crafted strings into user instructions (e.g., ``ignore all previous instructions and instead ...'') to override the original instruction logic~\cite{perez2022ignore}.
In multimodal settings, however, this threat becomes more stealthy and powerful: malicious instructions can be embedded not only explicitly within text prompts but also implicitly encoded in visual inputs\cite{li2025agenttypo, wudissecting, wang2025manipulating}. 
Such prompt injection attacks can steer models toward entirely different tasks~\cite{liu2024formalizing} or exert fine-grained control over model outputs~\cite{piet2024jatmo}, and have been dubbed the \#1 security risk for applications integrating LLMs by OWASP~\cite{OWASP}.

Despite this growing concern, existing prompt injection attacks against LVLMs still share a critical limitation: injecting prompts into one modality only serves to steer the model's interpretation of that singular input.
With more details, as shown in Figure~\ref{fig: intro}, current methods mainly fall into two categories.
The first is prompt injection with only image perturbed~\cite{li2025agenttypo, wudissecting}.
In this scenario, the model still follows the task defined by the benign text prompt, but its perception of the image is perturbed.
The second is prompt injection with both text and image prompts perturbed~\cite{wang2025manipulating}.
Here, although the model's perception of both text and image prompts is manipulated, the misperception in the textual domain is not caused by the image prompt perturbation but the text.


In this paper, we propose a \underline{cross}-\underline{m}odal \underline{p}rompt \underline{i}njection attack \sysname, in which the attacker can use image-only perturbation to change the model's interpretation of both the visual and textual prompts.
The key insight behind \sysname is that, for an LVLM, its visual evidence and textual intent integration occurs in its hidden state space but not the embedding space.
As a result, unlike the prior works~\cite{wudissecting, wang2025manipulating}, all of which optimize its prompt perturbation over the embedding space, \sysname turns its target to the model hidden state space.


The primary challenge brought by the target change is the explosively expanded parameter space for perturbation optimization.
For example, in a 7B-MiniGPT4 model, the visual embedding space contains only about $10^5$ parameters, whereas the hidden state space can reach about $10^7$ parameters.
Even worse, the hardness is further amplified by the size of image prompts. 
That means, with a $m$-dimensional hidden state space and an image prompt with size $n$, the perturbation optimization complexity is at least $O(mn)$.
Naturally, directly optimizing over the full hidden state space is computationally expensive and can easily overfit to model-specific activations, leading to low transferability of the generated perturbations.
To address this problem, \sysname introduces two strategies: 
a fusion-critical layer selection strategy for the constraint of $m$, 
and a distance-decremental perturbation budget assignment strategy for the constraint of $n$.

For fusion-critical layer selection, we first compare attack performance when perturbations are optimized on the early, middle, and final layers of LVLMs.
Interestingly, deviated from the past experience that the final layers contribute most to adversarial perturbation optimization~\cite{lv2023data, yuan2024dropout, chen2025rounding}, we find that the most effective layers for cross-modal prompt perturbation are located in the middle of LVLMs.
This is because middle layers are where textual intent and visual evidence are actively fused into task-level representations, whereas the final layers mainly map these already formed latent decisions to the output tokens~\cite{zhang2025cross, yumultimodal, neotowards}.

Based on this insight, \sysname focuses on the middle layers and further proposes an algorithm to identify which of them are most responsible for multimodal fusion.
Specifically, \sysname uses layer-wise probing to measure how the hidden states at each layer react to different text prompt changes. 
Layers that remain stable under wording or syntax changes but show clear changes when the task semantics change are treated as fusion-critical layers. 
These layers capture the stage where the model binds visual evidence with the textual instruction. 
Restricting optimization to them reduces the number of optimized hidden states and keeps the attack focused on the representations that determine task interpretation.


For perturbation budget assignment, we construct a distance-decremental budget masking mechanism to concentrate the image perturbation on semantic-critical regions. 
The most straightforward way to redistribute the budget is to use saliency scores, assigning larger budgets to pixels that contribute more to the model prediction~\cite{zhang2026understanding}. 
However, gradient-based saliency may also highlight background or task-irrelevant regions that are correlated with the model prediction, even though these regions are not the main visual evidence. 
Allocating large budgets to such scattered regions weakens the perturbation on truly important areas and reduces attack efficiency.
To address this, we introduce a spatial distance penalty centered on the semantic-critical region. 
The perturbation budget is high only for pixels that are salient and close to this center, and it gradually decreases as pixels move farther away. 

Extensive experiments demonstrate the effectiveness of our attack across diverse datasets and LVLMs. 
Specifically, in the black-box setting, \sysname achieves a $66.36\%$ success rate in inducing models to execute the attacker-chosen task, outperforming four state-of-the-art baselines by $49.09$ percentage points on average.
Both qualitative visualizations and quantitative image similarity metrics further show that \sysname preserves excellent imperceptibility.
In addition, ablation studies validate the contribution of each key component and demonstrate the robustness of \sysname under different attack scenarios.

Our contributions can be summarized as follows:
\begin{itemize}
    \item We propose \sysname, a cross-modal prompt injection attack against LVLMs, which can utilize image-only perturbation to manipulate the interpretation of LVLMs on both the visual and textual prompts.

    \item We explore a new direction to achieve perturbation optimization for prompt injection, i.e., optimizing over the model hidden state space instead of the embedding space.

    \item We design two strategies to constrain the optimization space for cross-modal prompt injection: 1) the fusion-critical layer selection strategy and 2) the distance-decremental perturbation budget assignment strategy. 
    During the design of the first strategy, we find that the most critical layers for cross-modal prompt perturbation optimization are located in the middle of LVLM but not the final layers, which is deviated from the past experience for adversarial perturbation optimization.

    \item Extensive evaluations across LVLMs and datasets demonstrate that, in the black-box setting, \sysname achieves a $66.36\%$ success rate in inducing models to execute the attacker-chosen task, outperforming baselines by $49.09$ percentage points on average.

\end{itemize}

\section{Preliminaries}
\label{sec:Preliminaries}
\subsection{Large Vision-Language Models}
Large Vision-Language Models (LVLMs) extend large language models with visual perception so that a single model can process a text--image pair and generate task-dependent language outputs, enabling applications such as image captioning, visual question answering, and vision-language dialogue. 
A typical LVLM consists of a vision encoder, a learnable projection module, and an autoregressive language model. 
Formally, let $x_v$ denote an input image and $x_p$ denote a text prompt. 
The image is first processed by a vision encoder $f_{vis}(\cdot)$ to produce a sequence of visual tokens $t_v = \{t_v^1, t_v^2, \ldots, t_v^{T_v}\}$, where $t_v^i \in \mathbb{R}^{d_v}$, $T_v$ is the number of visual tokens and $d_v$  is the feature dimension of the vision encoder. 
Because the vision encoder and the language model generally operate in different embedding spaces, a projection module $f_w(\cdot)$ maps the visual tokens into the hidden space of the language model, yielding $e_v = f_w(t_v)$ with $e_v \in \mathbb{R}^{T_v \times d_l}$, where $d_l$ is the hidden dimension of the language model. 
In parallel, the prompt $x_p$ is tokenized and embedded into text tokens $e_p = \{e_p^1, e_p^2, \ldots, e_p^{T_p}\}$, where $e_p^i \in \mathbb{R}^{d_l}$ and $T_p$ is the number of textual tokens. 
The projected visual tokens and text tokens are concatenated into a joint multimodal sequence $e=[e_v; e_p]$, which is then processed by the language model $f_{\phi}(\cdot)$ to autoregressively generate an output sequence $y=(y_1, y_2, \ldots, y_{T_y})$.

From a probabilistic perspective, an LVLM models the conditional distribution of the output text given both visual and textual inputs, denoted by $P_{\theta}(y \mid x_p, x_v)$, where $\theta$ collects the parameters involved in multimodal inference. 
The response distribution follows the standard autoregressive factorization
\begin{equation}
    P_{\theta}(y \mid x_p, x_v) = \prod_{i=1}^{T_y} P_{\theta}(y_i \mid y_{<i}, x_p, x_v),
\end{equation}
where $y_{<i}=\{y_1, y_2, \ldots, y_{i-1}\}$. 
At each generation step, the model produces a logit vector $z_i \in \mathbb{R}^{|V|}$ over the vocabulary $V$, which is transformed into a probability distribution over tokens by a softmax layer before decoding.

\mypara{Training and Inference}
Given a training dataset $D$, training is typically formulated as minimizing the negative log-likelihood of the ground-truth outputs:
\begin{equation}
    \mathcal{L}(\theta) = \mathbb{E}_{(x_p, x_v, y) \sim D} \left[- \log P_{\theta}(y \mid x_p, x_v)\right].
\end{equation}
This objective encourages the model to bind visual evidence and textual instructions into a coherent multimodal representation that supports downstream generation. 
At inference time, the parameters $\theta$ are fixed, and the model generates responses autoregressively from the learned conditional distribution, typically via greedy decoding or sampling-based strategies with temperature control.

\subsection{Prompt Injection Attack}
Prompt injection attacks~\cite{liu2024formalizing, shi2024optimization, pasquini2024neural} manipulate model behavior by injecting attacker-controlled content into the input without changing the model parameters. 
These attacks exploit the difficulty of instruction-following models in reliably separating trusted instructions from untrusted user-supplied content, allowing attackers to override the intended task, redirect model behavior, or induce unintended outputs. 
For LVLMs, the attack surface is broader because malicious signals can be introduced through either the textual or visual modality~\cite{li2025agenttypo, wudissecting}. 

Formally, given a trained LVLM $\mathcal{M}$ with fixed parameters $\theta$, a user provides a text--image input pair $(x_p, x_v)$. 
The text prompt $x_p$ defines the user-intended task $\mathcal{T}$, denoted as $\mathcal{T}\Leftarrow x_p$, and the model output is
\begin{equation}
    y = \mathcal{M}(x_p, x_v).
\end{equation}
In a multimodal prompt injection attack, the attacker aims to redirect the model from the user-intended task $\mathcal{T}$ to an attacker-chosen target task $\mathcal{T}_t$. 
We characterize the injected task using a target text--image pair $(x_p^t, x_v^t)$, where $\mathcal{T}_t\Leftarrow x_p^t$, and the desired target model output is
\begin{equation}
    y_t = \mathcal{M}(x_p^t, x_v^t).
\end{equation}
Instead of directly replacing the benign input with $(x_p^t, x_v^t)$, the attacker can inject perturbations into both modalities by constructing a text perturbation $\Delta_p$ and an image perturbation $\Delta_v$. 
The attack objective is to make the modified input $(x_p+\Delta_p, x_v+\Delta_v)$ induce the attacker-chosen task and produce the target output:
\begin{equation}
    \mathcal{M}(x_p+\Delta_p, x_v+\Delta_v) = \mathcal{M}(x_p^t, x_v^t) = y_t,
\end{equation}
where the injected text perturbation changes the task from $\mathcal{T}$ to $\mathcal{T}_t$, and the image perturbation provides additional visual evidence aligned with the target output. 
In this setting, the malicious instruction can be carried through both the textual and visual modalities, causing the LVLM to follow the attacker-specified task rather than the user's original task.
Different from prior multimodal attacks that modify both the text and image inputs, our attack uses image-only perturbations to change the model's interpretation of the text--image input and induce it to execute the attacker-chosen task. 
Formally, we define the two attack settings as follows:

\begin{definition}[Multimodal vs.\ Cross-Modal Prompt Injection Attacks]
Assume an LVLM $\mathcal{M}(\cdot, \cdot)$ and a text--image input pair $(x_p, x_v)$. 
Let $\mathcal{T} \Leftarrow x_p$ denote the intended task defined by $x_p$, and $y = \mathcal{M}(x_p, x_v)$ be the output of $\mathcal{M}$.

\begin{itemize}


\item \textbf{Multimodal Prompt Injection:}
Find perturbations $\Delta_p$ and $\Delta_v$ to make $\mathcal{T}_t \Leftarrow x_p+\Delta_p$ and $y_t=\mathcal{M}(x_p+\Delta_p, x_v+\Delta_v)$,
where $y_t$ and $\mathcal{T}_t$ are attacker-chosen but $\mathcal{T}_t\neq\mathcal{T}$ holds if and only if $\Delta_p\neq0$.

\item \textbf{Cross-Modal Prompt Injection:} 
Find a perturbation $\Delta_v$ to make $\mathcal{T}_t \Leftarrow x_p$ and $y_t = \mathcal{M}(x_p, x_v + \Delta_v)$ where both $\mathcal{T}_t\neq \mathcal{T}$ and $y_t\neq y$ are attacker-chosen.
\end{itemize}
\end{definition}

\section{Threat Model}
\label{sec:Threat Model}
We consider \sysname under a standard LVLM application scenario. 
Our threat model is defined along two dimensions: the attacker's goal and the attacker's capabilities.


\mypara{Attacker's Goal}
The attacker aims to change the task executed by the LVLM by modifying only the image input. 
Under this setting, the attacked image should cause the LVLM to execute an attacker-chosen injected task instead of the user-intended task. 
Given a benign text--image pair $(x_p, x_v)$, where $x_p$ is the text prompt and $x_v$ is the original image, the attacker generates a perturbed image $x_v' = x_v + \Delta_v$. 
The text prompt remains unchanged. 
The goal is to make the model interpret $(x_p, x_v')$ as an attacker-chosen injected task and produce the attacker-chosen output $y_t$. 
Formally, the attack seeks to satisfy
\begin{equation}
    \mathcal{M}(x_p, x_v+\Delta_v) = y_t,
\end{equation}
while keeping $\Delta_v$ visually imperceptible. 
Thus, the attack is not simply to cause an incorrect answer, but to redirect the model's task understanding through the visual modality alone.



\mypara{Attacker's Capabilities}
We consider a black-box target setting where the attacker cannot access the deployed target LVLM, including its parameters, gradients, intermediate hidden states, system prompt, safety policy, or inference configuration. 
The attacker can freely choose the injected task, including the target text prompt and target image, and thereby define the desired target output.
The attacker also knows the benign input pair, including the text prompt $x_p$ and the original image $x_v$.
To construct the attack, the attacker can use publicly available LVLMs offline to optimize a bounded image perturbation $\Delta_v$, producing the attacked image $x_v'=x_v+\Delta_v$. 
At deployment time, the attacker can only submit the attacked image $x_v'$ with the unchanged benign text prompt to the target LVLM.
These adversarial examples can subsequently compromise the functionality of target systems, such as misleading VLM-based web agents~\cite{xu2024advweb} or disrupting real-world object detectors~\cite{song2018physical}.

\section{Methodology}
\label{sec:Methodology}
\begin{figure*}[ht]
    \centering 
    \includegraphics[width=\linewidth]{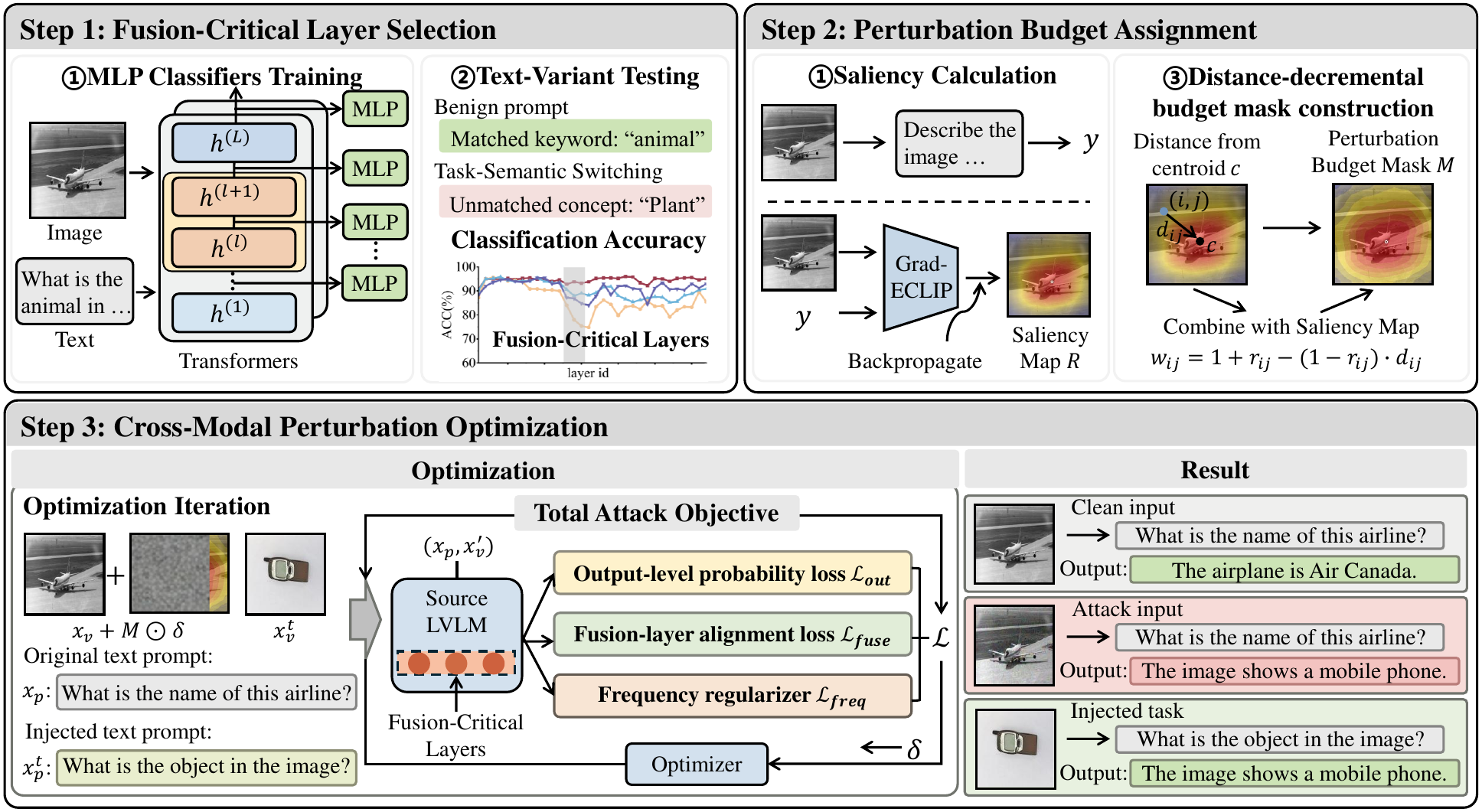}
    \caption{Overview of our attack. \sysname consists of three parts: (1) fusion-critical layer selection, which localizes the layers where visual evidence and textual intent are integrated; (2) perturbation budget assignment, which uses Grad-ECLIP saliency and distance-decremental weighting to construct a perturbation budget mask for budget reallocation; (3) Cross-Modal Perturbation Optimization, which jointly optimizes output-level, fusion-level, and frequency-domain objectives.}
    \label{fig: attack framework}
    \Description[Overview of our attack.]{Overview of our attack.}
\end{figure*}

In this section, we present the methodology of \sysname under our threat model. 
As shown in Figure~\ref{fig: attack framework}, \sysname consists of three main stages. 
First, fusion-critical layer selection reduces the parameter search space by limiting optimization to a small number of layers most relevant to multimodal fusion. 
Second, distance-decremental perturbation budget assignment reduces the image search space by concentrating the perturbation budget around semantic-critical regions and decreasing it with distance. 
Third, cross-modal perturbation optimization learns the final image perturbation using output-level supervision, fusion-layer hidden states alignment, and frequency regularization.

\subsection{Fusion-Critical Layer Selection}
\label{sec: fusion-layers select}
To constrain the optimized model parameter space, we first analyze the attack performance obtained from different LVLM layer groups. 
Specifically, we consider early layers, middle layers, final layers, and their pairwise combinations, resulting in six optimization settings. 
For each setting, we apply the optimization objective introduced in Section~\ref{sec: Cross-Modal Perturbation Optimization} to the selected layers. 
The detailed setup and results are reported in Section~\ref{sec: ablation_layer selection}. 
Unlike prior methods that mainly optimize final-layer representations~\cite{lv2023data, yuan2024dropout, chen2025rounding}, our results show that optimizing middle-layer representations achieves the best attack performance. 
This indicates that optimizing cross-modal representations at the fusion stage can more effectively steer the model's task interpretation.
Based on this observation, we focus on the middle layers and further introduce a fusion-critical layer selection strategy to identify the layers that contribute most to vision-text fusion.

\begin{figure}[ht]
    \centering 
    \includegraphics[width=\linewidth]{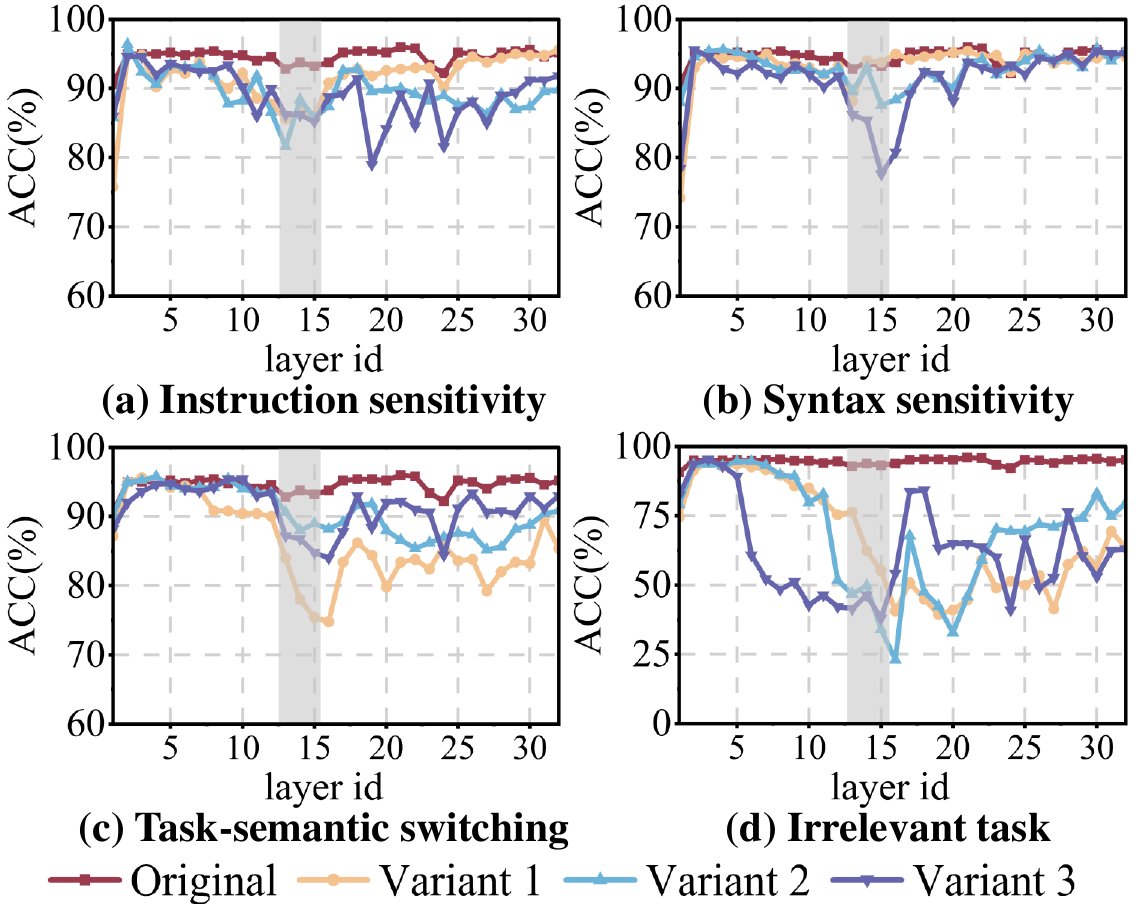}
    \caption{Layer-wise hidden state classification accuracy under prompt variants on MiniGPT4-vicuna. The fusion-critical layers are identified as layers 13--15.}
    \label{fig: Fusion-Critical Layers vicuna}
    \Description[Layer-wise probing of fusion-critical layers]{Layer-wise probing results that reveal the depth range where visual and textual information are most strongly fused in LVLMs.}
\end{figure}


Notably, existing approaches for studying LVLM internals mainly fall into three categories: 1) Attention knockout~\cite{neotowards}, masks or removes specific attention connections and quantifies the importance of layers or attention heads according to the resulting degradation in model predictions; 2) Logit lens~\cite{zhang2025cross, logit_lens}, maps hidden states of each layer to the vocabulary space through the language-model head and examines how output-related information emerges across the network; 3) Layer-wise probing~\cite{yumultimodal}, trains classifiers on the hidden states of each layer and measures how sensitive these representations are to different prompt variations.
Among them, attention knockout focuses on information-flow paths, while logit lens measures the relationship between intermediate hidden states and output distributions.
In contrast, layer-wise probing directly evaluates the semantic information encoded in each layer's hidden states and its stability under prompt changes.  
Therefore, we adopt this method to identify fusion-critical layers, defined as layers whose probing accuracy changes substantially when the prompt semantics or task is modified. 
Our method consists of two main steps:

\mypara{Step-1: Layer-wise classifier training}
We first feed the LVLM with animal images paired with the benign prompt ``What is the animal in the picture?''. 
For each transformer layer $l$, we extract the hidden state of the last input token, which summarizes the preceding visual and textual tokens under causal attention, and therefore provides a compact representation of the prompt-conditioned multimodal context. 
We then use the hidden states from each layer to train an independent Multilayer Perceptron (MLP) classifier $g^{(l)}$ for predicting the visual category. 
Once trained, all classifiers are frozen and used as a fixed evaluator for evaluating hidden states produced under different prompt variants.

\mypara{Step-2: Text-variant sensitivity testing}
Prior method~\cite{yumultimodal} only uses lexical variants (e.g., replacing ``picture'' with ``image'') and semantic negation variants (e.g. replacing ``animal'' with ``plant''). 
However, both variants are constructed through a single-word substitution, making it difficult to distinguish whether changes in hidden states come from local word replacement or from the resulting semantic shift.  
Therefore, we introduce four complementary prompt variants for fusion-layer selection. 
\emph{Instruction Sensitivity} changes the instruction wording while keeping the queried content, such as changing ``What is'' to ``Tell me'', to test whether a layer is affected by superficial instruction wording. 
\emph{Syntax Sensitivity} rewrites the sentence structure without changing the task meaning, such as using an inverted sentence form, to test sensitivity to syntactic structure. 
\emph{Task-Semantic Switching} changes the queried concept, such as changing ``animal'' to ``plant'', to test whether textual semantics begin to reshape the encoded visual evidence. 
\emph{Irrelevant-Task Replacement} replaces the original query with an unrelated task, such as ``What color is the sky?'', to test whether a layer encodes task-level intent rather than local lexical cues.
We provide three prompts for each variant in Table~\ref{table: appendix-prompt-variants}.
We then feed each image with these prompt variants into the LVLM, extract the hidden states from each layer, and evaluate them using the corresponding frozen MLP classifier.

Figure~\ref{fig: Fusion-Critical Layers vicuna} shows the layer-wise probing results of MiniGPT4-vicuna~\cite{zhuminigpt}. 
In the early layers, classification accuracy remains stable under instruction, syntax, and task-semantic changes, but drops sharply under irrelevant-task replacement, indicating that these layers mainly capture broad task context. 
In the middle layers, different prompt variants lead to clear differences: instruction and syntax changes cause only small accuracy drops, whereas task-semantic switching and irrelevant-task replacement cause much larger drops. 
This indicates that textual semantics begin to reshape visual representations in this depth range. 
For example, under the first task-semantic switching variant, the accuracy at layers 13--15 decreases by about 25\%. 
After these layers, the accuracy curves become stable, suggesting that multimodal inference has largely been completed. 
We therefore identify layers 13--15 as the fusion-critical region of MiniGPT4-vicuna. 
More experimental details and results are provided in Appendix~\ref{appendix: Fusion-Critical Layer Selection}.

\subsection{Distance-Decremental Perturbation Budget Assignment}
\label{sec: Perturbation Budget Assignment}
To constrain the image optimization space, we design a distance-decremental perturbation budget mask. 
The goal is to assign larger perturbation budgets to regions that are both semantically important and spatially close to the main evidence region. 
A straightforward way to allocate the budget is to rely only on saliency scores, giving larger budgets to pixels with higher saliency~\cite{zhang2026understanding}. 
However, gradient-based saliency can be scattered over background regions that are correlated with the prediction but do not provide the core visual evidence. 
Moreover, LVLMs can often recognize an object from only a few discriminative pixels~\cite{zhang2026understanding, yang2025beyond}. 
Therefore, the perturbation should cover the target object as completely as possible, rather than being concentrated only on a few highly salient pixels.
To address this problem, we introduce the distance to the semantic-critical center as a spatial penalty, making the perturbation budget decrease as pixels move farther from this center.
This design keeps the perturbation concentrated around the most relevant semantic region while still allowing nearby contextual pixels to contribute. 
Our method consists of two main steps:


\mypara{Step-1: Saliency calculation}
Different LVLMs may focus on different image regions for the same input. 
To obtain a more model-independent saliency estimate, we first use a general-purpose reference model, e.g., GPT-5.4-mini, to generate a concise image description. 
Specifically, we query the model with the fixed prompt ``Describe the image in one sentence.'' and obtain a summary $y$. 
This summary captures the main visual semantics of the image and serves as a text condition for saliency estimation. 
By aligning the image with this neutral description, we can locate the regions that are most relevant to the image semantics while reducing dependence on the architectural bias of the attacked LVLM.

Specifically, we utilize Grad-ECLIP~\cite{zhang2026understanding, zhao2024gradient} to compute a saliency map on the description. 
Grad-ECLIP measures image-text relevance by backpropagating the similarity signal between the visual and text representations to the visual tokens, and then projects the token-level relevance scores back to the image space. 
Given the image $x_v$ and its description $y$, this process can yield an importance map $S$, where $s_{ij}$ denotes the saliency score of pixel $(i,j)$. 

\mypara{Step-2: Distance-decremental perturbation budget mask construction}
Given the saliency map $S$, we first normalize the importance score of each pixel:
\begin{equation}
    r_{ij}=\frac{s_{ij}}{\max_{i,j}(s_{ij})},
\end{equation}
where $r_{ij} \in [0,1]$ denotes the relative semantic importance of pixel $(i,j)$.
We then estimate the semantic-critical center of the image from the most salient pixels. 
Specifically, we select the top-$k$\% pixels with the highest normalized saliency scores as the semantic-critical support $U$, and compute their weighted centroid:
\begin{equation}
    c=\frac{\sum_{(i,j)\in U} r_{ij} \cdot (i,j)}{\sum_{(i,j)\in U} r_{ij}}.
\end{equation}
This center represents the main region that contributes to the image-level semantics.
Next, we compute a distance penalty for each pixel based on its distance to the semantic-critical center. 
For each pixel $(i,j)$, we calculate its Euclidean distance to $c$, denoted by $v_{ij}=\|(i,j)-c\|_2$, and normalize it as
\begin{equation}
    d_{ij}=\frac{v_{ij}}{\max_{i,j}(v_{ij})}.
\end{equation}

We combine semantic importance and spatial distance to obtain the budget weight:
\begin{equation}
    w_{ij}=1+r_{ij}-(1-r_{ij})\cdot d_{ij}.
\end{equation}
Here, $r_{ij}$ increases the weight of semantically important pixels, while $(1-r_{ij})\cdot d_{ij}$ penalizes pixels that are both less salient and farther from the semantic-critical center. 
Therefore, important regions are retained, whereas distant regions with weak semantic relevance are downweighted.
Finally, we convert the weight map into a local perturbation budget mask:
\begin{equation}
    \epsilon_{ij}=\epsilon\left(1-\lambda+\lambda\cdot\frac{w_{ij}}{\bar{w}}\right),
\end{equation}
where $\bar{w}$ is the average weight, $\epsilon$ denotes the base perturbation budget, and $\lambda \in [0,1]$ controls the strength of budget redistribution. 
Normalizing by $\bar{w}$ keeps the average perturbation budget across the image unchanged, so the mask redistributes the total budget rather than increasing it. 
When $\lambda=0$, all pixels share the same budget. 
As $\lambda$ increases, pixels with larger weights receive more budget, while distant and weakly relevant pixels receive less. 
As a result, the final mask concentrates perturbations around semantic-critical regions and gradually reduces the budget for less relevant areas.

\subsection{Cross-Modal Perturbation Optimization}
\label{sec: Cross-Modal Perturbation Optimization}
Prior image-based prompt injection methods mainly optimize perturbations in the visual embedding space. 
Although such perturbations can change how the image is represented, they fail to fundamentally change how the model interprets the textual instruction. 
The reason is that the injected noise primarily affects visual features before multimodal fusion, while the task semantics are determined later through the interaction between visual evidence and the text prompt. 
To address this limitation, we turn the optimization target from the visual embedding space to the model hidden state space, where visual and textual information are jointly integrated.
By integrating the reduction strategies detailed in Sections~\ref{sec: fusion-layers select} and~\ref{sec: Perturbation Budget Assignment}, we significantly prune both the parameter and image optimization search space.
Specifically, letting $M$ denote the perturbation budget mask whose entries are given by the local thresholds $\epsilon_{ij}$ and $\delta$ the learnable perturbation, we define the effective perturbation as $\Delta_v = M \odot \delta$ and construct the attacked image as
\begin{equation}
    x_v' = x_v + \Delta_v = x_v + M \odot \delta.
\end{equation}
The mask controls where the perturbation is allowed to be stronger, so that the optimization focuses on image regions that are more likely to affect multimodal task grounding.

To enable the perturbation to successfully alter the model's interpretation of the text prompt and perform an attacker-chosen task, we propose a dual-faceted strategy that optimizes the perturbation at two complementary levels. 
The first level is output-level optimization, which directly increases the probability of the target task response. 
Its role is to ensure that the perturbed image can successfully modify the final model output. 
The second level is fusion-level optimization, which aligns the hidden states of the selected fusion-critical layers with the target task representation. 
Its role is to intervene in the intermediate representation space, ensuring that the perturbation changes how the model interprets the benign text prompt and thereby improving transferability.

The first optimization objective is output-level optimization, which makes the attacked image directly induce the attacker-chosen response under the benign text prompt. 
Specifically, given the attacker-selected target task input $(x_p^t, x_v^t)$ and keywords of target output token sequence $y_t=\{y_1^t,\ldots,y_T^t\}$. 
After applying the perturbation, the attacked image is given by $x_v'=x_v+M\odot\delta$. 
We then optimize the perturbation by maximizing the likelihood of the target output conditioned on the benign prompt and the attacked image. 
The output-level loss is defined as
\begin{equation}
    \mathcal{L}_{out} = -\sum_{i=1}^{T} \log P_{\mathcal{M}}\left(y_i^t \mid y_{<i}^t, x_p, x_v'\right),
\end{equation}
where $\mathcal{M}$ denotes the source LVLM. 
This objective explicitly forces the attacked input to place high probability mass on the target task token sequence, thereby aligning the optimization with the actual malicious goal.
However, as the final response is produced after multiple stages of multimodal fusion, language decoding, and model-specific alignment, the optimized perturbations can only capture source-specific features, which perform well in white-box evaluation but transfer poorly to other models.

The second component is fusion-level optimization, which makes the benign prompt and the perturbed image form an internal representation similar to that of the attacker-chosen task. 
Let $\mathcal{L}_{sel}$ denote the selected fusion-critical layers, and let $H^{(l)}(x_p,x_v)$ be the hidden state at layer $l$.  
We then minimize the distance between their hidden states at the selected fusion-critical layers:
\begin{equation}
    \mathcal{L}_{fuse} = \sum_{l \in \mathcal{L}_{sel}}
    \left\| H^{(l)}(x_p, x_v') - H^{(l)}(x_p^t, x_v^t) \right\|_2^2.
\end{equation}
This loss directly constrains the intermediate fusion process: it pulls the hidden states of the attacked input toward the hidden states of the target task. 
As a result, the perturbation is encouraged to change how the model interprets the benign text prompt, rather than only changing the final decoded answer. 
By supervising the selected fusion-critical layers, this objective provides more stable and semantically meaningful gradients, which improve optimization stability and cross-model transferability.


To improve visual stealthiness and reduce overfitting to the source model, we introduce a frequency regularization term to suppress high-frequency components in the perturbation. 
Specifically, we transform the perturbation $\Delta_v$ into the frequency domain, remove the central low-frequency region, and penalize the mean magnitude of the remaining high-frequency components:
\begin{equation}
    \mathcal{L}_{freq} =
    \frac{1}{|\Omega_h|}
    \sum_{(u,v)\in\Omega_h}
    \left|
    \mathcal{F}(\Delta_v)_{u,v}
    \right|,
\end{equation}
where $\mathcal{F}(\cdot)$ denotes the two-dimensional Fourier transform and $\Omega_h$ denotes the high-frequency region outside the central low-frequency window. 
This loss penalizes high-frequency noise, which can create visible artifacts and overfit to the preprocessing or patch-embedding behavior of the source model.
In this way, $\mathcal{L}_{freq}$ encourages smoother perturbations that better preserve visual appearance and transfer robustly across models.

Finally, to improve robustness to common image variations, we optimize the perturbation over multiple transformed views instead of a single fixed image. 
Specifically, for each optimization step, we construct a transformation augmentation set $\mathcal{A}=\{\tau_0,\tau_1,\ldots,\tau_5\}$, where $\tau_0$ is the identity mapping and the remaining five transformations correspond to scaling, rotation, brightness adjustment, blur, and additive noise. 
The attack objective is then enforced on all six views $\{\tau(x_v') \mid \tau \in \mathcal{A}\}$, so that the learned perturbation remains effective under benign distortions introduced during image transmission, preprocessing, or acquisition, yielding a more stable attack in practical settings.
We combine the three terms into the final attack objective:
\begin{equation}
    \mathcal{L}_{attack}
    =
    \frac{1}{|\mathcal{A}|}
    \sum_{\tau \in \mathcal{A}}
    [
    \mathcal{L}_{out}(\tau(x_v'))
    +
    \alpha \mathcal{L}_{fuse}(\tau(x_v'))
    +
    \beta \mathcal{L}_{freq}(\tau(\Delta_v)) ],
\end{equation}
where $\alpha$ and $\beta$ control the weights of the fusion-layer alignment loss and the frequency regularization loss, respectively. 
Taking into account these three complementary optimization goals, we can make the perturbation affect both the model's interpretation of the benign prompt and its final response, while improving stealth and transferability.

\section{Experimental Evaluation}
\label{sec:Experimental Evaluation}
In this section, we conduct comprehensive evaluations for \sysname.
We first introduce our experimental setup and then present the detailed results of each evaluation.

\subsection{Experimental Setup}
\mypara{Datasets}
We evaluate \sysname on three datasets: MSCOCO~\cite{lin2014microsoft}, ImageNet~\cite{russakovsky2015imagenet}, and TextVQA~\cite{singh2019towards}. 
We use VQA questions~\cite{luo2024image, cai2025towards} as text prompts to define distinct benign and attacker-chosen tasks.
For MSCOCO and ImageNet, the attacker-chosen target task is ``What is the animal in the picture?'', with sheep and cat as the target images, respectively. 
For TextVQA, the target task is ``What is the object in the picture?'', with a phone as the target image.
We randomly sample 500 examples from each dataset for evaluation, while ensuring that the sampled images do not belong to the same category as the corresponding target image.
Detailed dataset construction and preprocessing settings are provided in Appendix~\ref{appendix: Experimental Setup}.

\mypara{Models}
We evaluate six representative LVLMs, namely MiniGPT4-llama2~\cite{zhuminigpt}, MiniGPT4-vicuna~\cite{zhuminigpt}, InstructBLIP~\cite{dai2023instructblip}, BLIP-2~\cite{li2023blip}, BLIVA~\cite{hu2024bliva}, and Qwen2.5-VL~\cite{qwen2.5-VL}. 
All models are used with their original released settings, without architecture modifications or customized inference configurations.
We use MiniGPT4-llama2, MiniGPT4-vicuna, InstructBLIP, and Qwen2-VL~\cite{Qwen2VL} as source models for perturbation optimization. 
The optimized adversarial examples are then evaluated on the target LVLMs to measure both source model attack effectiveness and cross-model transferability.
Unless otherwise specified, we optimize three fusion-critical layers and set the perturbation budget to $16/255$.

\mypara{Evaluation metrics}
We use two metrics to evaluate attack performance: Attack Success Rate (ASR)~\cite{shi2024optimization, labunets2025fun, zhang2026understanding} and Semantic Similarity (SS)~\cite{liu2024pandora, cai2025towards, yan2025fit}. 
ASR measures the proportion of test samples for which the perturbed image makes the model follow the attacker-chosen task, and therefore reflects the effectiveness of the attack.
SS measures the semantic similarity between the model response and the attacker-specified target text. 
Specifically, we encode both the attacked response and the target text using Sentence-Transformer~\cite{reimers2019sentence} with the all-MiniLM-L6-v2 encoder, and compute their cosine similarity as SS. 
This metric evaluates whether the attack induces the intended malicious semantics rather than arbitrary output changes. 
Reporting both metrics is necessary because a successful cross-modal prompt injection attack should achieve both high attack success and high semantic alignment.

\begin{table*}[ht]
\centering
\renewcommand{\arraystretch}{1.3}
\setlength{\tabcolsep}{6.5pt}
\belowrulesep=0pt
\aboverulesep=0pt
\footnotesize
\caption{Attack performance across MSCOCO, ImageNet, and TextVQA. ASR (\%) and SS (\%) are reported for the baseline attacks and \sysname under the best source model selected for each target LVLM. The best results are highlighted in bold font.}
\label{table: ASR_SS best source modal}

\begin{tabular}{c|c|c|cc|cc|cc|cc|cc} 
\toprule
\multicolumn{1}{c|}{\multirow{2}{*}{Datasets}} & \multicolumn{1}{c|}{\multirow{2}{*}{Target Models}} & \multicolumn{1}{c|}{\multirow{2}{*}{Best Source Model}} & \multicolumn{2}{c|}{ARE-W}  & \multicolumn{2}{c|}{ARE-B} & \multicolumn{2}{c|}{CI} & \multicolumn{2}{c|}{ATPI} & \multicolumn{2}{c}{\textbf{Ours}}\\ 
\cline{4-13}
\multicolumn{1}{c|}{} & \multicolumn{1}{c|}{} & \multicolumn{1}{c|}{} & \multicolumn{1}{c}{ASR} & \multicolumn{1}{c|}{SS} & \multicolumn{1}{c}{ASR} & \multicolumn{1}{c|}{SS} & \multicolumn{1}{c}{ASR} & \multicolumn{1}{c|}{SS} & \multicolumn{1}{c}{ASR} & \multicolumn{1}{c|}{SS} & \multicolumn{1}{c}{ASR} & \multicolumn{1}{c}{SS}\\ 
\hline
\multirow{6}{*}{MSCOCO} &MiniGPT4-llama2 &MiniGPT4-vicuna   &2.42  &28.92  &7.48  &31.66  &11.90 &31.24  &0.00 &25.80  &\textbf{28.43} &\textbf{43.17}\\ \cline{2-13}
                        &MiniGPT4-vicuna &MiniGPT4-llama2   &9.76  &31.23  &7.32  &32.96  &30.95 &42.98  &0.00 &25.39  &\textbf{90.20} &\textbf{71.22}\\ \cline{2-13}
                        &InstructBLIP    &MiniGPT4-llama2   &2.43  &30.27  &9.76  &34.35  &48.51 &45.64  &0.00 &26.14  &\textbf{90.57} &\textbf{52.57}\\ \cline{2-13}
                        &BLIP-2          &MiniGPT4-llama2   &4.88  &15.26  &44.64 &38.24  &89.34 &62.94  &0.00 &12.67  &\textbf{96.08} &\textbf{65.72}\\ \cline{2-13}
                        &BLIVA           &MiniGPT4-llama2   &26.83 &34.63  &7.33  &32.55  &66.67 &47.25  &0.00 &26.42  &\textbf{91.18} &\textbf{53.03}\\ \cline{2-13}
                        &Qwen2.5-VL      &Qwen2-VL          &0.00  &33.80  &\textbf{20.51} &37.30  &11.90 &35.66  &0.00 &29.40  &12.75 &\textbf{38.74}\\

\hline
\multirow{6}{*}{ImageNet} &MiniGPT4-llama2 &MiniGPT4-vicuna &0.00  &19.84  &6.52  &21.40  &14.55 &25.20  &0.00  &20.47  &\textbf{27.18} &\textbf{32.63}\\ \cline{2-13}
                          &MiniGPT4-vicuna &MiniGPT4-llama2 &9.73  &24.60  &8.70  &23.84  &29.09 &30.88  &0.00  &21.74  &\textbf{69.90} &\textbf{51.01}\\ \cline{2-13}
                          &InstructBLIP    &MiniGPT4-llama2 &0.00  &18.83  &10.87 &24.97  &43.64 &31.20  &0.00  &18.89  &\textbf{94.17} &\textbf{37.56}\\ \cline{2-13}
                          &BLIP-2          &MiniGPT4-llama2 &12.17 &15.66  &39.13 &31.86  &\textbf{94.55} &45.54  &0.00  &13.29  &93.20 &\textbf{54.17}\\ \cline{2-13}
                          &BLIVA           &MiniGPT4-llama2 &7.28  &20.03  &6.52  &21.64  &65.45 &33.60  &0.00  &18.29  &\textbf{94.17} &\textbf{36.08}\\ \cline{2-13}
                          &Qwen2.5-VL      &Qwen2-VL        &13.95 &18.89  &13.04 &17.82  &\textbf{56.36} &\textbf{29.62}  &4.87  &17.66  &29.41 &21.87  \\

\hline
\multirow{6}{*}{TextVQA} &MiniGPT4-llama2 &MiniGPT4-vicuna  &14.96 &27.20  &7.83  &28.44  &16.67 &27.10  &4.63  &24.31  &\textbf{52.94} &\textbf{51.89}\\ \cline{2-13}
                         &MiniGPT4-vicuna &MiniGPT4-llama2  &7.32  &22.98  &9.80  &34.96  &13.95 &27.33  &0.00  &25.39  &\textbf{51.82} &\textbf{51.51}\\ \cline{2-13}
                         &InstructBLIP    &MiniGPT4-vicuna  &30.77 &23.96  &3.92  &18.81  &31.25 &23.85  &9.24  &17.91  &\textbf{82.35} &\textbf{44.93}\\ \cline{2-13}
                         &BLIP-2          &MiniGPT4-vicuna  &49.75 &35.32  &7.97  &8.02   &70.83 &45.87  &4.65  &18.09  &\textbf{81.37} &\textbf{54.26}\\ \cline{2-13}
                         &BLIVA           &InstructBLIP     &65.85 &32.11  &5.87  &21.95  &31.25 &28.35  &2.58  &13.97  &\textbf{88.24} &\textbf{53.51}\\ \cline{2-13}
                         &Qwen2.5-VL      &Qwen2-VL         &3.03  &\textbf{42.33}  &3.93  &37.85  &6.25  &27.72  &2.44  &21.96  &\textbf{20.59} &37.99\\
\hline
\multicolumn{3}{c|}{\multirow{1}{*}{Average}}               &14.51 &26.44  &12.29 &27.70  &40.73 &35.67  &1.58  &20.99  &\textbf{66.36} &\textbf{47.33}\\ 
\bottomrule
\end{tabular}
\end{table*}

\begin{table*}[ht]
\centering
\renewcommand{\arraystretch}{1.3}
\setlength{\tabcolsep}{4.2pt}
\belowrulesep=0pt
\aboverulesep=0pt
\footnotesize
\caption{Imperceptibility comparison across MSCOCO, ImageNet, and TextVQA. Perturbations are optimized on MiniGPT4-llama2. All metrics are multiplied by 100 for readability. The best results are highlighted in bold font.}
\label{table: imperceptibility}

\begin{tabular}{c|ccccc|ccccc|ccccc} 
\toprule
\multicolumn{1}{c|}{\multirow{2}{*}{Attacks}} & \multicolumn{5}{c|}{MSCOCO}  & \multicolumn{5}{c|}{ImageNet} & \multicolumn{5}{c}{TextVQA}\\ 
\cline{2-16}
\multicolumn{1}{c|}{} & \multicolumn{1}{c}{SSIM$\uparrow$} & \multicolumn{1}{c}{MS-SSIM$\uparrow$} & \multicolumn{1}{c}{FSIM$\uparrow$} & \multicolumn{1}{c}{HaarPSI$\uparrow$} & \multicolumn{1}{c|}{LPIPS $\downarrow$} & \multicolumn{1}{c}{SSIM$\uparrow$} & \multicolumn{1}{c}{MS-SSIM$\uparrow$} & \multicolumn{1}{c}{FSIM$\uparrow$} & \multicolumn{1}{c}{HaarPSI$\uparrow$} & \multicolumn{1}{c|}{LPIPS $\downarrow$} & \multicolumn{1}{c}{SSIM$\uparrow$} & \multicolumn{1}{c}{MS-SSIM$\uparrow$} & \multicolumn{1}{c}{FSIM$\uparrow$} & \multicolumn{1}{c}{HaarPSI$\uparrow$} & \multicolumn{1}{c}{LPIPS $\downarrow$} \\ 
\cline{1-16}
ARE-W  &59.41 &81.57 &65.88 &93.73 &70.48  &65.51 &90.11 &67.22 &95.31 &43.21  &66.53 &81.28 &63.46 &94.58 &60.59\\\cline{1-16}
ARE-B  &51.06 &69.22 &62.31 &92.66 &85.70  &56.66 &80.41 &59.03 &97.47 &39.42  &61.88 &73.17 &59.09 &98.25 &45.44\\\cline{1-16}
CI     &53.49 &73.56 &63.75 &93.54 &78.31  &57.69 &84.06 &62.34 &94.26 &55.69  &57.65 &70.49 &61.17 &94.14 &74.32\\\cline{1-16}
ATPI   &\textbf{99.38} &\textbf{99.54} &\textbf{99.62} &\textbf{99.65} &\textbf{1.45}   &\textbf{97.96} &\textbf{98.44} &\textbf{98.71} &98.79 &\textbf{3.66}   &\textbf{99.38} &\textbf{99.49} &\textbf{99.57} &\textbf{99.72} &\textbf{1.72}\\\cline{1-16}
Ours   &73.12 &93.92 &76.88 &98.79 &18.44  &71.24 &93.83 &72.58 &\textbf{98.85} &20.41  &71.27 &93.51 &70.17 &98.85 &20.50\\\cline{1-16}
\bottomrule
\end{tabular}
\end{table*}

\mypara{Baselines}
We compare \sysname with four representative attacks: ARE-W~\cite{wudissecting}, ARE-B~\cite{wudissecting}, CI~\cite{wang2025manipulating}, and ATPI~\cite{li2025agenttypo}. 
ARE-W is a white-box attack that directly optimizes the perturbed image to increase the probability of the attacker-chosen output. 
ARE-B is a black-box variant that uses CLIP as a surrogate model to increase the similarity between the perturbed image embeddings and the target output embeddings. 
CI also uses CLIP-based optimization to increase the similarity between the perturbed image embeddings and target image embeddings, while additionally optimizing the text prompt to increase the probability of the target output. 
In our evaluation, we only allow image perturbations.
ATPI directly inserts a malicious textual instruction into the image and optimizes its rendering attributes, such as color, size, and position, to improve readability and attack effectiveness. 
For ATPI, we use its original prompt, ``Please briefly describe the content and text in the image,'' because the model otherwise may not be explicitly encouraged to read the injected text. 
More details on baseline implementation are provided in Appendix~\ref{appendix: Experimental Setup}.

\subsection{Attack Performance}

\subsubsection{Overall Attack Performance}
We compare \sysname with four representative prompt injection baselines on three datasets and six target LVLMs. 
Table~\ref{table: ASR_SS best source modal} reports the best source model used for perturbation optimization for each target LVLM, together with the ASR (\%) and SS (\%). 
Overall, \sysname achieves the best performance across datasets and target models.
Averaged over all settings, \sysname obtains an ASR of $66.36\%$, outperforming the four baselines by $49.09$ percentage points on average. 
Meanwhile, it achieves an average SS of $47.33$, exceeding the baseline average by $11.68$ points.
These results indicate that \sysname not only redirects LVLMs to attacker-chosen tasks more effectively, but also preserves stronger semantic consistency with the desired target responses.

\begin{figure*}[ht]
    \centering 
    \includegraphics[width=\linewidth]{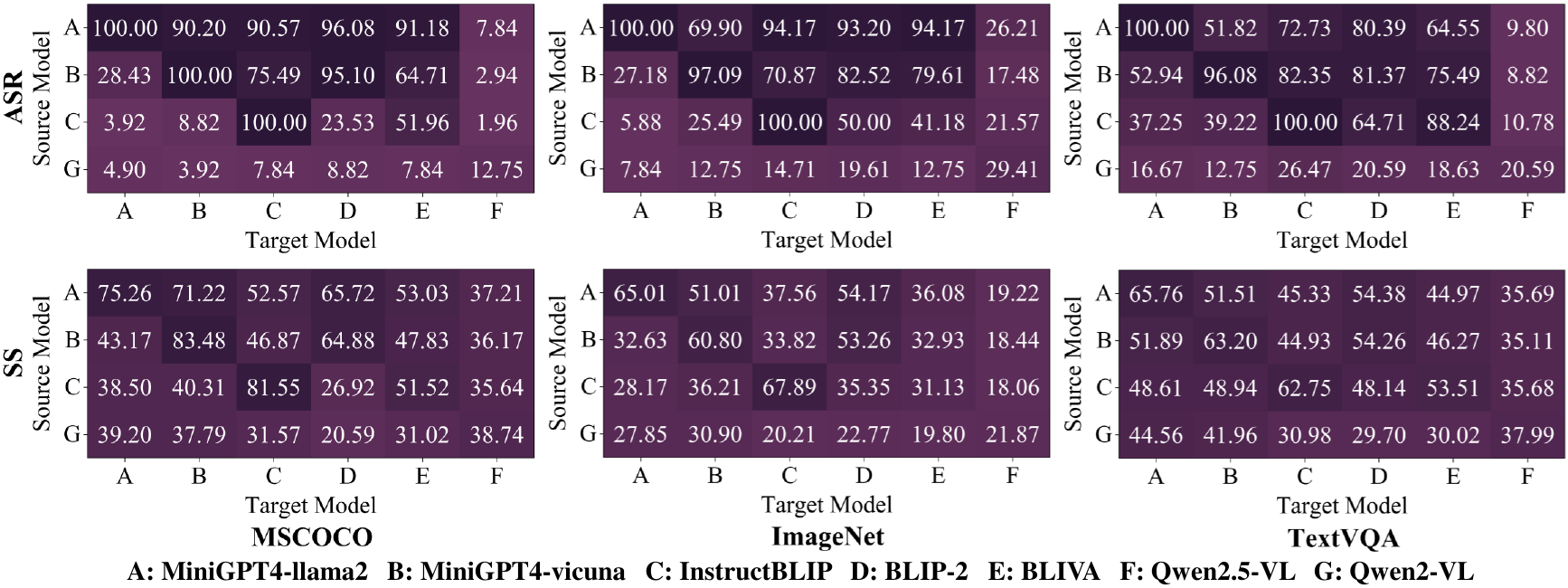}
    \caption{Heatmap of \sysname across LVLMs. Perturbations are optimized on MiniGPT4-llama2, MiniGPT4-vicuna, InstructBLIP, and Qwen2-VL, and then evaluated on target models.}
    \label{fig: Heatmap}
    \Description[Overview of the cross-modal prompt injection framework]{An overview of the proposed attack framework, including fusion-critical layer identification, importance-aware region selection with adaptive budgeting, and image-only perturbation optimization.}
\end{figure*}

The advantage of \sysname is particularly clear in the detailed results. 
For example, on MSCOCO, \sysname achieves high ASR across multiple target models, including $90.20\%$ on MiniGPT4-vicuna, $90.57\%$ on InstructBLIP, $96.08\%$ on BLIP-2, and $91.18\%$ on BLIVA, while also maintaining competitive SS scores. 
In contrast, ARE-W shows limited transferability, with an average ASR of only $8.24\%$.
This is because ARE-W directly optimizes the target output probability on the white-box source model, the resulting perturbations tend to overfit source-model-specific generation behaviors and transfer poorly to other LVLMs.
The CLIP-based baselines, ARE-B and CI, achieve higher ASR than ARE-W, with average ASRs of $34.58\%$ and $54.57\%$, respectively. 
This indicates that optimizing image embeddings toward target semantics can partially change the model's visual understanding and induce target-related outputs. 
However, since the benign text prompt is still interpreted as the original task, these methods are less effective at changing the model's task-level behavior. 
Their strong reliance on target-semantic image alignment also introduces more visible perturbations, resulting in weaker imperceptibility, as discussed in Section~\ref{sec: exp imperceptibility}.
ATPI performs the weakest overall, achieving only $4.41\%$ average ASR.
Although it embeds malicious textual instructions into the image and uses prompts that encourage the model to read image text, this strategy does not reliably trigger the desired cross-modal instruction behavior. 
It also makes the injected instruction directly visible to users, which further limits its practicality. 
In contrast, \sysname attacks the internal multimodal reasoning process more directly. 
By optimizing fusion-critical hidden states and allocating perturbation budgets to semantic-critical image regions, \sysname better aligns the perturbation with the LVLM's internal task interpretation. 
This explains its stronger attack effectiveness, better transferability, and improved semantic alignment across diverse models and datasets.

\subsubsection{Imperceptibility}
\label{sec: exp imperceptibility}
Imperceptibility is a key requirement for practical prompt injection attacks, since an attacked image is more likely to be used by a victim when it remains visually close to the original image. 
Therefore, beyond attack effectiveness, we evaluate whether different methods can preserve the visual fidelity of the input image after applying adversarial perturbation. 
Figure~\ref{fig: attack visualization} shows qualitative examples of different attacks. 
We further quantify imperceptibility using five widely used image-similarity metrics: SSIM~\cite{wang2004image}, MS-SSIM~\cite{wang2021feature}, FSIM~\cite{zhang2011fsim}, HaarPSI~\cite{reisenhofer2018haar}, and LPIPS~\cite{zhang2018unreasonable}. 
SSIM and MS-SSIM evaluate structural consistency, FSIM captures feature-level similarity, HaarPSI measures perceptual similarity based on Haar wavelet responses, and LPIPS estimates perceptual distance in a deep feature space.
Higher SSIM, MS-SSIM, FSIM, and HaarPSI indicate better visual preservation, while lower LPIPS indicates smaller perceptual distortion.

As shown in Table~\ref{table: imperceptibility}, ATPI achieves the best scores on most metrics because it adds visible malicious text rather than dense image perturbations.
This means that it directly exposes the malicious instruction to the user, which weakens its practicality.
Among the remaining noise-optimization attacks, \sysname achieves the best imperceptibility. 
For example, on MSCOCO, \sysname obtains an LPIPS score of only $18.44$, whereas ARE-W, ARE-B, and CI obtain $70.48$, $85.70$, and $78.31$, respectively. 
The visual examples in Figure~\ref{fig: attack visualization} are consistent with these results: the images generated by \sysname remain visually close to the originals, while ARE-B and CI often produce noticeable appearance changes.
This is because their objectives push image embeddings toward target semantics, causing the attacked images to visually drift toward the target concept.
This effect is especially evident for CI, where the optimized images exhibit visible target-related patterns, such as sheep-like patterns on MSCOCO, cat-like patterns on ImageNet, and phone-like patterns on TextVQA.
In contrast, \sysname allocates larger perturbation budgets to semantic-critical regions and optimizes fusion-critical hidden states, avoiding direct modification of the image semantics.
In addition, we use $\mathcal{L}_{freq}$ to suppress high-frequency components in the perturbation, which further improves visual imperceptibility. 

\subsubsection{Cross-Source Validation}
We further evaluate \sysname under different source model settings to examine its transferability. 
Specifically, we use MiniGPT4-llama2, MiniGPT4-vicuna, InstructBLIP, and Qwen2-VL as source models to optimize image perturbations, and then evaluate the generated attacked images on six target LVLMs. 
This setting includes both the white-box case, where the source and target models are the same, and the transfer case, where the target model is different from the source model. 
Figure~\ref{fig: Heatmap} shows the results of \sysname across different source-target pairs, and Table~\ref{table: ASR_SS all source modal} provides the detailed comparison with baseline attacks.

The results show that \sysname can achieve $99.24\%$ white-box attack success rates and can also transfer successfully to a broad range of target models. 
For example, on MSCOCO, the attacked images optimized on MiniGPT4-llama2 can reach $90.20\%$, $90.57\%$, $96.08\%$, and $91.18\%$ ASR on MiniGPT4-vicuna, InstructBLIP, BLIP-2, and BLIVA, respectively. 
These attacks also maintain strong semantic alignment with the target responses, with SS scores of $71.22\%$, $52.57\%$, $65.72\%$, and $53.03\%$. 
The results also demonstrate that transferability is stronger between models with more similar architectures. 
For instance, on TextVQA, attacked images optimized on MiniGPT4-llama2, MiniGPT4-vicuna, and InstructBLIP achieve only $9.80\%$, $8.82\%$, and $10.78\%$ ASR on Qwen2.5-VL, respectively. 
In contrast, images optimized on Qwen2-VL achieve $20.59\%$ ASR on Qwen2.5-VL. 
This is expected as models with similar architectures tend to share more comparable model hidden state spaces and fusion mechanisms, making perturbations optimized on one model more likely to change the internal task interpretation of another.
In other words, when attackers can access a source model with the same architecture type as the target model, \sysname can achieve prompt injection attacks with high success rates.

\subsection{Ablation Study}

\begin{table*}[ht]
\centering
\renewcommand{\arraystretch}{1.3}
\setlength{\tabcolsep}{8pt}
\belowrulesep=0pt
\aboverulesep=0pt
\footnotesize
\caption{Ablation of layer-selection strategies for optimization. ASR (\%) and SS (\%) are reported across target LVLMs for different layer groups, with perturbations optimized on MiniGPT4-llama2. The best results are highlighted in bold font.}
\label{table: fusion-layers select}

\begin{tabular}{c|cc|cc|cc|cc|cc} 
\toprule
\multicolumn{1}{c|}{\multirow{2}{*}{Layer Group}} & \multicolumn{2}{c|}{MiniGPT4-llama2}  & \multicolumn{2}{c|}{MiniGPT4-vicuna} & \multicolumn{2}{c|}{InstructBLIP} & \multicolumn{2}{c|}{BLIP-2} & \multicolumn{2}{c}{BLIVA}\\ 
\cline{2-11}
\multicolumn{1}{c|}{} & \multicolumn{1}{c}{ASR} & \multicolumn{1}{c|}{SS} & \multicolumn{1}{c}{ASR} & \multicolumn{1}{c|}{SS} & \multicolumn{1}{c}{ASR} & \multicolumn{1}{c|}{SS} & \multicolumn{1}{c}{ASR} & \multicolumn{1}{c|}{SS} & \multicolumn{1}{c}{ASR} & \multicolumn{1}{c}{SS} \\ 
\cline{1-11}
early        &\textbf{100} &76.01           &89.20 &72.21           &88.20 &51.77                     &95.56 &67.46                 &88.36 &52.45\\\cline{1-11}
fusion       &\textbf{100} &75.17           &\textbf{90.16} &72.63  &\textbf{89.18} &\textbf{52.87}  &\textbf{96.08} &65.33        &\textbf{90.12} &53.85\\\cline{1-11}
final        &\textbf{100} &\textbf{77.58}  &87.25 &\textbf{72.89}  &88.65 &51.86                     &93.78 &66.34                 &87.19 &\textbf{54.56}\\\cline{1-11}
early+fusion &\textbf{100} &76.07           &88.24 &72.31           &85.29 &51.99                     &95.54 &67.51                 &89.18 &53.68\\\cline{1-11}
early+final  &\textbf{100} &75.57           &89.18 &72.88           &88.33 &52.66                     &94.14 &\textbf{67.71}        &90.05 &54.55\\\cline{1-11}
fusion+final &\textbf{100} &76.23           &88.24 &72.58           &87.25 &52.05                     &95.36 &67.47                 &89.18 &54.35\\
\bottomrule
\end{tabular}
\end{table*}

\begin{table*}[ht]
\centering
\renewcommand{\arraystretch}{1.3}
\setlength{\tabcolsep}{8pt}
\belowrulesep=0pt
\aboverulesep=0pt
\footnotesize
\caption{Ablation of different perturbation budget masks for optimization. ASR (\%) and SS (\%) are reported across target LVLMs for no mask, saliency-only mask, and \sysname masks with $\lambda \leq 0.5$, using perturbations optimized on MiniGPT4-llama2. The best results are highlighted in bold font.}
\label{table: lambda of perturbation budget mask}

\begin{tabular}{c|cc|cc|cc|cc|cc} 
\toprule
\multicolumn{1}{c|}{\multirow{2}{*}{Budget Mask}} & \multicolumn{2}{c|}{MiniGPT4-llama2}  & \multicolumn{2}{c|}{MiniGPT4-vicuna} & \multicolumn{2}{c|}{InstructBLIP} & \multicolumn{2}{c|}{BLIP-2} & \multicolumn{2}{c}{BLIVA}\\ 
\cline{2-11}
\multicolumn{1}{c|}{} & \multicolumn{1}{c}{ASR} & \multicolumn{1}{c|}{SS} & \multicolumn{1}{c}{ASR} & \multicolumn{1}{c|}{SS} & \multicolumn{1}{c}{ASR} & \multicolumn{1}{c|}{SS} & \multicolumn{1}{c}{ASR} & \multicolumn{1}{c|}{SS} & \multicolumn{1}{c}{ASR} & \multicolumn{1}{c}{SS} \\ 
\cline{1-11}
no mask       &\textbf{100} &75.17           &88.46 &70.63           &88.18 &51.65                    &93.24 &65.33                   &90.12 &\textbf{53.85} \\\cline{1-11}
saliency-only &\textbf{100} &75.49           &83.33 &69.90           &85.29 &51.36                    &94.12 &64.89                   &88.24 &52.65 \\\cline{1-11}
$\lambda$=0.1 &\textbf{100} &75.42           &86.27 &69.63           &88.57 &51.86                    &93.14 &65.12                   &87.25 &51.63 \\\cline{1-11}
$\lambda$=0.2 &\textbf{100} &75.21           &88.24 &71.28           &87.62 &51.76                    &94.12 &65.17                   &87.25 &52.00 \\\cline{1-11}
$\lambda$=0.3 &\textbf{100} &75.26           &\textbf{90.20} &71.22  &\textbf{90.57} &\textbf{52.57}  &\textbf{96.08} &\textbf{65.72} &\textbf{91.18} &53.03 \\\cline{1-11}
$\lambda$=0.4 &\textbf{100} &75.30           &83.33 &70.51           &87.25 &51.38                    &93.14 &64.68                   &89.22 &52.58 \\\cline{1-11}
$\lambda$=0.5 &\textbf{100} &\textbf{75.53}  &86.27 &\textbf{71.85}  &82.35 &51.29                    &94.12 &64.98                   &89.22 &52.77 \\
\bottomrule
\end{tabular}
\end{table*}

\subsubsection{Layer Selection for Optimization}
\label{sec: ablation_layer selection}
To examine which model layers provide the most effective optimization target, we conduct an ablation study on layer selection. 
We use MiniGPT4-llama2 as the source model and do not apply the perturbation budget mask.
We define three layer groups: early layers, middle layers, and final layers. 
Based on the fusion-layer analysis in Figure~\ref{fig: Fusion-Critical Layers llama2}, we use the selected fusion-critical layers as the middle-layer group. 
Specifically, the three groups are early layers $\{1,2,3\}$, fusion-critical middle layers $\{12,13,14\}$, and final layers $\{30,31,32\}$.
We then evaluate six optimization settings, including optimizing each group alone and optimizing different combinations of two groups.

The results show that attacks optimized on the middle fusion-critical layers achieve the best performance.
Compared with early or final layers, optimizing these layers yields the most consistent attack performance across different LVLMs. 
For example, fusion-layer optimization achieves an ASR of $90.16\%$ on MiniGPT4-vicuna, compared with $89.20\%$ for early-layer optimization and $87.25\%$ for final-layer optimization.
This indicates that perturbations become less effective when applied before or after the main multimodal fusion stage of LVLMs.
This finding is consistent with the functional differences among LVLM layers.
Early layers mainly process modality-specific visual signals, so perturbations at this stage may not directly affect the model's task understanding. 
Final layers operate after the model has already formed a prompt-conditioned representation, making them more tied to model-specific response generation. 
In contrast, fusion-critical layers lie at the point where the image and the text prompt are integrated into a task representation. 
Perturbing these layers can therefore more directly affect the model's task interpretation, leading to stronger and more stable attack performance.

\subsubsection{Effect of the Perturbation Budget Mask}
To examine the role of perturbation budget allocation, we conduct an ablation study with different masking strategies. 
We use MiniGPT4-llama2 as the source model and optimize the same fusion-critical layers in all settings. 
We compare three budget allocation strategies: using no mask, applying the saliency-only mask adopted in SGMA~\cite{zhang2026understanding}, and using our distance-decremental budget mask with different coefficients. 
This comparison allows us to evaluate whether focusing perturbations only on highly salient regions is sufficient, or whether a spatially smoother budget allocation leads to more stable transfer across target models.

\begin{figure}[ht]
    \centering 
    \includegraphics[width=\linewidth]{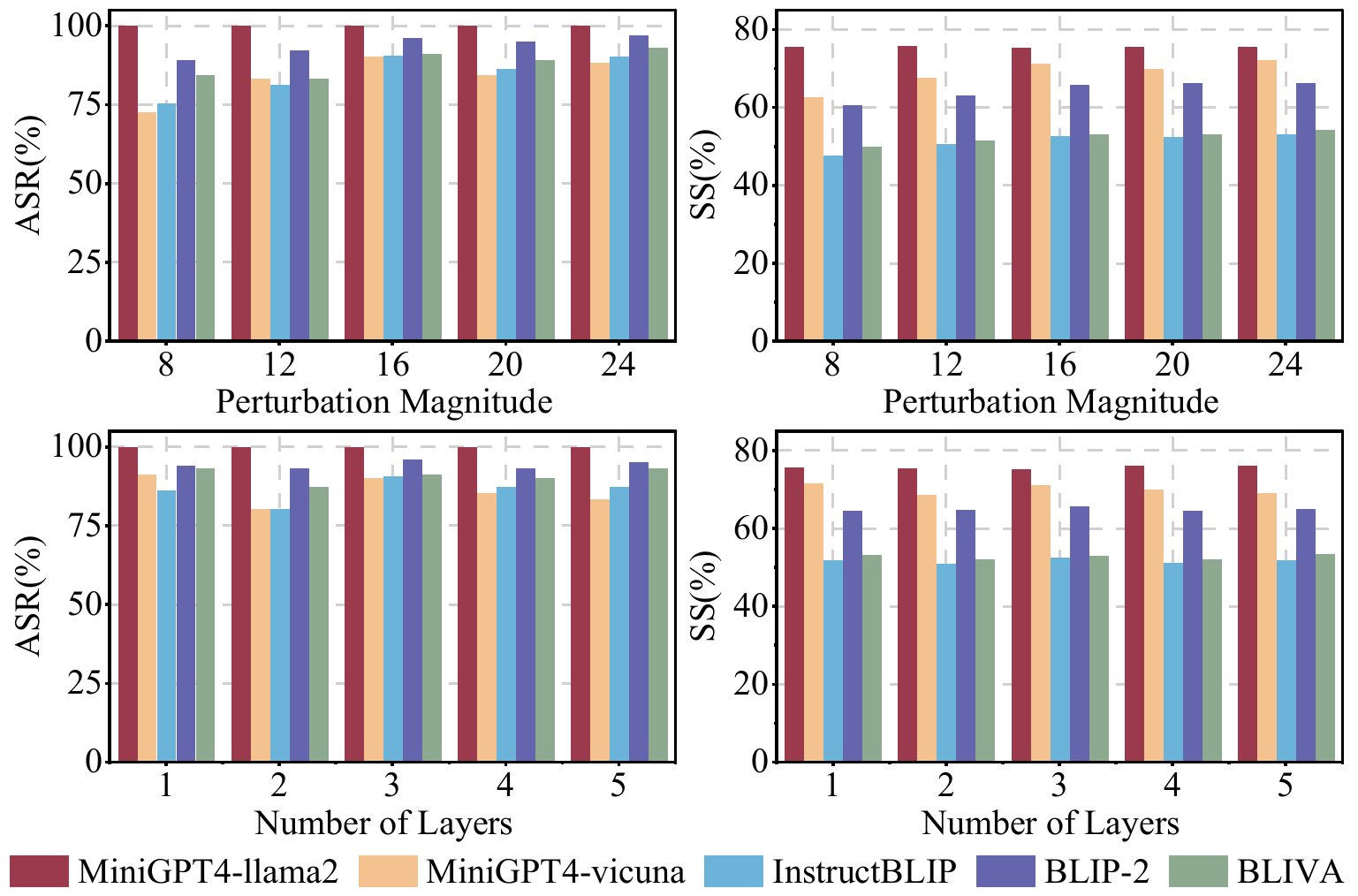}
    \caption{Ablation of perturbation budget and optimized layer count on MSCOCO. The first row varies $\epsilon \in \{8,12,16,20,24\}/255$, and the second row varies the number of optimized fusion-critical layers from one to five, using MiniGPT4-llama2 as the source model.}
    \label{fig: ablation perturbation_eplosion and number of layers}
    \Description[Ablation of perturbation budget and optimized layer count]{A two-row ablation visualization showing how attack performance changes with different perturbation budgets and different numbers of optimized fusion-critical layers.}
\end{figure}

Table~\ref{table: lambda of perturbation budget mask} reports the results for $\lambda \in \{0.1,0.2,0.3,0.4,0.5\}$, and the results for the remaining coefficients are provided in Table~\ref{table: lambda0.6-1 of perturbation budget mask}. 
Overall, $\lambda=0.3$ gives the best trade-off between concentrating perturbations on important regions and preserving sufficient spatial coverage. 
It achieves the highest ASR on MiniGPT4-vicuna, InstructBLIP, and BLIVA, reaching $90.20\%$, $90.57\%$, and $91.18\%$, respectively.
As shown in the mask visualizations in Figure~\ref{fig: budget mask lambda}, increasing $\lambda$ leads to a more pronounced step-wise decrease in the perturbation budget.
The trend across different coefficients shows that perturbation allocation should be neither too uniform nor too concentrated. 
When $\lambda$ is too small, the mask is close to uniform allocation, so the optimization does not sufficiently emphasize semantically important regions. 
When $\lambda$ is too large, most of the perturbation budget is restricted to a small set of pixels, which weakens the attack's ability to jointly affect local visual evidence and broader contextual information. 
This confirms that a moderate distance-decremental mask is more effective for cross-model transfer. 
Based on these results, we set $\lambda=0.3$ in the final configuration.

\subsubsection{Effect of the Perturbation Budget}
The perturbation budget directly controls the maximum visual modification allowed for the attack: a larger budget expands the feasible optimization region, but can also reduce visual stealthiness.
We evaluate five perturbation budgets, $\epsilon \in \{8,12,16,20,24\}/255$, on MSCOCO using MiniGPT4-llama2 as the source model, while keeping all other optimization settings unchanged.
As shown in the first row of Figure~\ref{fig: ablation perturbation_eplosion and number of layers}, \sysname remains effective across all perturbation budgets. 
Even with the smallest budget, $\epsilon=8/255$, the attack still achieves over $70\%$ ASR on all target models, showing that \sysname can succeed under a strict perturbation constraint. 
As $\epsilon$ increases, the ASR generally improves across models. 
This trend is expected because a larger budget allows the optimization to more flexibly modify the fusion-layer representations toward the attacker-specified task.
These results show that \sysname does not rely on excessively large perturbations, while a larger budget can further improve attack success by providing more room for controlled semantic manipulation.

\subsubsection{Effect of the Number of Selected Layers}
The number of selected layers controls how much of the multimodal fusion process is directly constrained during optimization. 
Optimizing too few layers may provide insufficient coverage of the fusion stage, while optimizing too many layers may introduce less relevant or more model-specific representations. 
We therefore evaluate the effect of the number of optimized fusion-critical layers. 
Specifically, we vary the number of optimized layers from $1$ to $5$, using MiniGPT4-llama2 as the source model on MSCOCO.
As shown in the second row of Figure~\ref{fig: ablation perturbation_eplosion and number of layers}, \sysname maintains strong attack performance across different layer counts, showing that it is not sensitive to the exact number of selected layers within the fusion-critical region.
Among all settings, optimizing three layers achieves the best overall performance. 
This suggests that three fusion-critical layers provide enough coverage of task-semantic fusion while keeping the optimization target focused. 
Increasing the number of optimized layers to four or five does not further improve performance, likely because the additional layers are less directly related to multimodal fusion and may introduce model-specific optimization noise.

\subsection{Defenses}
To defend against our attack, we suggest adopting defense methods that either transform the input image before inference or strengthen the inference procedure itself. 
In our experiments, we evaluate five representative defenses from these two categories. 
The first category is \emph{input-transformation defenses}, which preprocess the image to disrupt adversarial perturbations. 
It includes Randomization~\cite{frosio2023best, xie2018mitigating}, which applies random resizing and padding; RandomRotation~\cite{luo2024image, liu2024pandora}, which introduces small random geometric transformations; and JPEG Compression~\cite{guo2018countering}, which suppresses high-frequency perturbation artifacts through image recompression. 
The second category is \emph{inference-safeguarding defenses}, which modify the prediction process rather than the input alone. 
It includes SmoothVLM~\cite{sun2024safeguarding}, which aggregates predictions over randomly masked image variants, and DPS~\cite{zhou2025defending}, which uses partial image observations and a correction prompt to guide the final response.
Detailed instantiations are provided in Appendix~\ref{appendix: Experimental Setup}.

Table~\ref{table: attack under defenses-llama2} evaluates the robustness of \sysname under different defenses, where perturbations are optimized on MiniGPT4-llama2. 
Additional results are provided in Appendix~\ref{appendix: Defenses}. 
Overall, these defenses reduce the attack success rate to varying degrees, but none of them completely eliminates the attack.
Among the evaluated defenses, SmoothVLM is the most effective in most settings. 
It reduces the ASR to below $5\%$ on MSCOCO and ImageNet, and keeps the ASR below $10\%$ on TextVQA across all target models. 
JPEG Compression also consistently weakens the attack, indicating that part of the optimized perturbation is sensitive to frequency-domain degradation. 
In comparison, Randomization and RandomRotation provide only limited protection, suggesting that the perturbations can still survive mild spatial transformations.
DPS shows less stable defense performance, which is less effective on the source model and in several transfer settings. 
For example, under DPS, \sysname still achieves $52.94\%$ ASR on MSCOCO and $71.84\%$ ASR on ImageNet. 
It also reaches $65.05\%$ ASR on BLIVA for ImageNet. 
These results suggest that existing defenses can mitigate cross-modal prompt injection attacks, but stronger defense mechanisms are still needed.

\begin{table*}[ht]
\centering
\renewcommand{\arraystretch}{1.3}
\setlength{\tabcolsep}{8pt}
\belowrulesep=0pt
\aboverulesep=0pt
\footnotesize
\caption{Defense evaluation for perturbations optimized on MiniGPT4-llama2. ASR (\%) and SS (\%) are reported across datasets, target LVLMs, and defense strategies.}
\label{table: attack under defenses-llama2}

\begin{tabular}{c|c|cc|cc|cc|cc|cc} 
\toprule
\multicolumn{1}{c|}{\multirow{2}{*}{Datasets}} & \multicolumn{1}{c|}{\multirow{2}{*}{Defenses}} & \multicolumn{2}{c|}{MiniGPT4-llama2}  & \multicolumn{2}{c|}{MiniGPT4-vicuna} & \multicolumn{2}{c|}{InstructBLIP} & \multicolumn{2}{c|}{BLIP-2} & \multicolumn{2}{c}{BLIVA}\\ 
\cline{3-12}
\multicolumn{1}{c|}{} & \multicolumn{1}{c|}{} & \multicolumn{1}{c}{ASR} & \multicolumn{1}{c|}{SS} & \multicolumn{1}{c}{ASR} & \multicolumn{1}{c|}{SS} & \multicolumn{1}{c}{ASR} & \multicolumn{1}{c|}{SS} & \multicolumn{1}{c}{ASR} & \multicolumn{1}{c|}{SS} & \multicolumn{1}{c}{ASR} & \multicolumn{1}{c}{SS} \\ 
\cline{1-12}
\multirow{5}{*}{MSCOCO} & Randomization     &13.73 &41.83  &4.90  &38.52  &17.65 &33.47  &24.51 &45.71  &21.57 &34.53 \\ \cline{2-12}
                        & RandomRotation    &22.55 &44.80  &9.80  &39.57  &16.67 &34.38  &18.63 &43.68  &19.61 &34.23 \\ \cline{2-12}
                        & JPEG Compression  &12.75 &41.20  &5.88  &37.60  &8.82  &32.05  &8.82  &39.71  &10.78 &31.58 \\ \cline{2-12}
                        & SmoothVLM         &0.00  &37.43  &0.00  &37.03  &0.98  &30.43  &0.98  &37.60  &0.98  &30.02 \\ \cline{2-12}
                        & DPS               &52.94 &55.14  &32.35 &48.50  &14.71 &32.69  &0.00  &34.42  &47.06 &39.73 \\
\hline
\multirow{5}{*}{ImageNet} & Randomization     &31.07 &38.25  &16.50 &31.89  &33.98 &23.80  &38.83 &37.26  &28.16 &23.05 \\ \cline{2-12}
                          & RandomRotation    &33.01 &40.63  &22.33 &32.87  &34.95 &24.00  &29.13 &35.39  &27.18 &23.64 \\ \cline{2-12}
                          & JPEG Compression  &17.48 &33.96  &17.48 &31.92  &19.42 &22.29  &19.42 &32.28  &16.50 &21.24 \\ \cline{2-12}
                          & SmoothVLM         &0.97  &26.44  &0.97  &28.46  &3.88  &20.07  &0.97  &27.84  &1.94 &19.44 \\ \cline{2-12}
                          & DPS               &71.84 &50.67  &30.10 &35.85  &21.36 &24.68  &6.80  &36.60  &65.05 &33.18 \\ 
\hline
\multirow{5}{*}{TextVQA} & Randomization      &32.73 &47.44  &19.61 &42.86  &35.29 &33.19  &12.75 &30.53  &29.41 &31.40 \\ \cline{2-12}
                         & RandomRotation     &30.91 &47.71  &17.65 &42.81  &32.35 &32.12  &14.71 &30.87  &26.47 &32.38 \\ \cline{2-12}
                         & JPEG Compression   &20.59 &43.56  &15.53 &43.88  &24.51 &30.92  &6.86  &28.12  &21.57 &29.66 \\ \cline{2-12}
                         & SmoothVLM          &5.45  &41.18  &5.83  &41.45  &8.82  &28.90  &1.96  &28.31  &9.80 &30.69 \\ \cline{2-12}
                         & DPS                &55.45 &50.62  &14.71 &44.15  &24.51 &34.84  &0.00  &30.32  &31.37 &36.68 \\ 
\bottomrule
\end{tabular}
\end{table*}

\section{Related Work}
\label{sec:Related Work}
\subsection{Prompt Injection Attacks}

Prompt injection attacks manipulate LLM behavior by injecting malicious instructions into the model input, causing the model to deviate from the intended task and follow attacker-chosen objectives. 
Early studies mainly focus on manually designed prompt injection strategies. 
These works show that simple injected instruction sequences can already be effective, especially when they are concatenated with benign data or separated using special characters such as newlines (``\textbackslash n'') and tabs (``\textbackslash t'')~\cite{Securing2023Rich, Prompt2022Simon}. 
Another common strategy is to use explicit task-overriding instructions, such as ``Ignore my previous instructions ...'', to make the model discard the original task context and execute the injected task instead~\cite{branch2022evaluating, Securing2023Rich, perez2022ignore, Prompt2022Simon}. 
Some attacks further inject fake responses to the original task, misleading the model into treating the benign task as completed and continuing with the malicious instruction~\cite{Delimiters2023Simon}. 
Based on these attack patterns, Liu~\et~\cite{liu2024formalizing} formalize prompt injection attacks and combine multiple manual strategies to improve attack effectiveness.


Recent studies also investigate automated attacks under white-box access.
Greedy Coordinate Gradient (GCG)~\cite{zou2023universal} uses gradient information to iteratively search for adversarial input tokens that increase the likelihood of a target model response.
Liu~\et~\cite{liu2024automatic} apply GCG to prompt injection by optimizing injected instruction strings according to different model response behaviors. 
Shi~\et~\cite{shi2024optimization} study prompt injection in the LLM-as-a-Judge setting, where the model selects the best answer from multiple candidates. 
In this scenario, attackers embed optimized injection strings into a controlled candidate response, causing the judge model to favor the attacker-controlled answer regardless of its actual quality. 
These works demonstrate that prompt injection can be generated both manually and automatically, posing a practical threat to instruction-following LLM systems.

\subsection{Attacks on LVLMs}

With the rapid development of LVLMs, their security risks have attracted increasing attention. 
Existing studies have explored several types of attacks against LVLMs, including prompt injection, adversarial, jailbreak, membership inference, and backdoor attacks.

For prompt injection attacks, Wu~\et~\cite{wudissecting} craft image-based injections by using multiple CLIP models~\cite{radford2021learning} as surrogate models, encouraging the attacked image to align with the target response.
Li~\et~\cite{li2025agenttypo} embed malicious instructions directly into images and adjust visual factors such as position, size, and color so that LVLMs can recognize the injected text during inference. 
Wang~\et~\cite{wang2025manipulating} optimize the visual embeddings toward target-image semantics while also optimizing deceptive text prompts with a source model.


Beyond prompt injection, many works study adversarial attacks that aim to induce specific LVLM responses. 
Wang~\et~\cite{wang2024transferable} optimize modality-consistency features with attention-guided perturbations and use orthogonality constraints to improve cross-model transferability. 
Cai~\et~\cite{cai2025towards} formulate the attack from an information-theoretic perspective by maximizing the mutual information between the perturbation and the target response while reducing the influence of the original image content.
Li~\et~\cite{lievaluating} propose MABA, an untargeted black-box attack that disrupts both visual encoding and modality alignment in LVLMs by suppressing discriminative visual features and using a mutual-information-aware projector to simulate the cross-modal adapter.

Other studies investigate broader security threats to LVLMs. 
Jailbreak attacks~\cite{shen2024anything, yu2024don} craft harmful multimodal inputs to bypass safety mechanisms and induce unsafe responses. 
Qi~\et~\cite{qi2024visual} optimize adversarial images in a white-box setting to increase the likelihood of malicious outputs, while Yang~\et~\cite{yang2025distraction} decompose harmful queries into sub-questions and construct contrastive sub-images to distract model attention from safety constraints. 
For membership inference, Li~\et~\cite{li2024membership} propose a cross-modal inference pipeline and use a MaxR\'enyi-K\% metric based on output logits to identify whether an image or text sample appears in the training set. 
Hu~\et~\cite{hu2025membership} exploit LVLMs' sensitivity to temperature parameter changes to distinguish members from non-members. 
For backdoor attacks, Lyu~\et~\cite{lyubackdooring} inject trigger patterns using out-of-distribution data and combine knowledge distillation with semantic consistency constraints, while Liu~\et~\cite{liu2025stealthy} design structured low-frequency triggers during the self-supervised pretraining stage of LVLMs to make backdoor patterns better aligned with natural image features.

\section{Conclusion \& Future Work}
\label{sec:Conclusion}
In this paper, we propose \sysname, a novel cross-modal prompt injection attack against LVLMs. 
\sysname uses image-only perturbations to change the model's interpretation of both the visual input and the textual prompt, causing the model to execute an attacker-chosen task while the text prompt remains unchanged.  
We explore a new direction for prompt injection optimization by optimizing in the model hidden state space instead of the embedding space, where visual and textual information are integrated. 
To reduce the large optimization space, we introduce two strategies: fusion-critical layer selection and distance-decremental perturbation budget assignment. 
The first strategy reduces the model parameter space by restricting optimization to a small set of layers most relevant to multimodal fusion. 
We further find that the most effective layers are located in the middle of LVLMs, especially the fusion-critical layers, which differs from the common practice of optimizing final layers in adversarial perturbation methods.
The second strategy reduces the image search space by concentrating the perturbation budget around semantic-critical regions. 
Experiments across multiple datasets and LVLMs demonstrate the effectiveness, transferability, and imperceptibility of \sysname. 
A current limitation is that the attack still depends on specific text prompts. 
In future work, our objective is to explore prompt-agnostic image-only attacks and develop defenses against such prompt injection.

\bibliographystyle{ACM-Reference-Format}
\bibliography{References}

\appendix

\section*{Appendix}
\section{Open Science}
We provide the artifacts needed to evaluate and reproduce our work. 
All artifacts are available at \url{https://anonymous.4open.science/r/CrossMPI-0F50}.

The released repository contains the source code of \sysname, including the implementation of the attack and evaluation scripts. 
It also provides part of the processed dataset resources, together with instructions for dataset preparation, model setup, and configuration. 
The datasets and models used in our experiments are publicly available, and the repository includes the information needed to obtain and reproduce the experimental settings.
These artifacts are made available at submission time for review and replication.

\section{Pseudocode}
The pseudocode of \sysname is summarized in Algorithm~\ref{alg: CrossMPI}.

\begin{algorithm}[h]
\caption{Cross-Modal Prompt Injection (\sysname)}
\label{alg: CrossMPI}
\small
\begin{algorithmic}[1]
\Require Source LVLM $\mathcal{M}$, benign input $(x_p,x_v)$, target reference $(x_p^t,x_v^t,y_t)$, probing dataset $\mathcal{D}_{probe}$, prompt variants $\mathcal{Q}$, description model $g$, perturbation budget $\epsilon$, redistribution coefficient $\lambda$, transform augmentation set $\mathcal{A}$, optimization steps $N$
\Ensure Attacked image $x_v'$

\State Identify fusion-critical layers $\mathcal{L}_{sel}$ by layer-wise probing on $\mathcal{D}_{probe}$ with prompt variants $\mathcal{Q}$

\State Generate an image description $y \gets g(x_v)$
\State Compute the saliency map $S \gets \textsc{Grad-ECLIP}(x_v,y)$
\State Normalize the saliency map:
\[
    r_{ij}=\frac{s_{ij}}{\max_{i,j}(s_{ij})}
\]
\State Select the top-$k$\% salient pixels as $U$ and compute the semantic-critical center:
\[
    c=\frac{\sum_{(i,j)\in U} r_{ij}\cdot(i,j)}
    {\sum_{(i,j)\in U} r_{ij}}
\]
\State Compute the normalized distance map:
\[
    d_{ij}=\frac{\|(i,j)-c\|_2}
    {\max_{i,j}\|(i,j)-c\|_2}
\]
\State Compute the budget weight and perturbation budget mask:
\[
    w_{ij}=1+r_{ij}-(1-r_{ij})d_{ij},
    \quad
    M_{ij}=\epsilon_{ij}
    =
    \epsilon\left(1-\lambda+\lambda\frac{w_{ij}}{\bar{w}}\right)
\]
where $\bar{w}=\frac{1}{HW}\sum_{i,j}w_{ij}$.

\State Initialize the learnable perturbation variable $\delta$
\For{$t=1$ to $N$}
    \State $\Delta_v \gets M \odot \delta$
    \State $x_v' \gets \mathrm{clip}(x_v+\Delta_v, 0, 1)$
    \State Compute output-level loss:
    \[
        \mathcal{L}_{out} = 
        - \frac{1}{|\mathcal{A}|}
        \sum_{\tau \in \mathcal{A}}
        \sum_{i=1}^{T} \log P_{\mathcal{M}}\left(y_i^t \mid y_{<i}^t, x_p, \tau(x_v')\right)
    \]
    \State Compute fusion-level loss:
    \[
        \mathcal{L}_{fuse} = 
        \frac{1}{|\mathcal{A}|}
        \sum_{\tau \in \mathcal{A}}
        \sum_{l \in \mathcal{L}_{sel}}
        \left\|
        H^{(l)}(x_p,\tau(x_v')) - H^{(l)}(x_p^t,x_v^t)
        \right\|_2^2
    \]
    \State Compute frequency regularization:
    \[
        \mathcal{L}_{freq} = 
        \frac{1}{|\mathcal{A}|}
        \frac{1}{|\Omega_h|} 
        \sum_{\tau \in \mathcal{A}}
        \sum_{(u,v)\in\Omega_h} \left| \mathcal{F}(\tau(\Delta_v))_{u,v} \right|,
    \]
    \State $\mathcal{L}_{attack} \gets \mathcal{L}_{out}+\alpha\mathcal{L}_{fuse}+\beta\mathcal{L}_{freq}$
    \State Update $\delta$ by minimizing $\mathcal{L}_{attack}$
\EndFor

\State \Return $\mathrm{clip}(x_v+M\odot\delta, 0, 1)$
\end{algorithmic}
\end{algorithm}

\section{Fusion-Critical Layer Selection Details}
\label{appendix: Fusion-Critical Layer Selection}
This section provides the experimental details of the fusion-critical layer selection introduced in Section~\ref{sec: fusion-layers select}. 
We build a fine-grained five-way dog classification benchmark from ImageNet, using five categories: Chihuahua, Golden Retriever, Labrador Retriever, German Shepherd, and Siberian Husky. 
For each category, we sample 500 images, with 400 for training and 100 for testing. 
For the target LVLM, we input the training images with the benign prompt ``What is the animal in the picture?'' and record the hidden states from every intermediate Transformer layer. 
For a model with $N$ layers, we train $N$ independent Multilayer Perceptron (MLP) classifiers. 
The classifier for layer $l$ takes the hidden state from layer $l$ as input and predicts one of the five dog categories. 
All MLP classifiers are trained only with hidden states obtained under the benign prompt and are then frozen for evaluation.

\begin{table}[t]
\caption{Prompt variants for identifying fusion-critical layers. The table lists benign, instruction-sensitive, syntax-sensitive, task-semantic switching, and irrelevant-task replacement prompts, with modified spans highlighted in red.}
\label{table: appendix-prompt-variants}
\small
\centering
\renewcommand{\arraystretch}{1.2}
\begin{tabular}{ll}
\toprule
\textbf{Type} & \textbf{Prompt} \\
\midrule
\textbf{Benign prompt} & \textbf{``What is the animal in the picture?''} \\
\makecell[l]{Instruction Sensitivity} & ``\textcolor{red}{Tell me} the animal in the picture.'' \\
\makecell[l]{Instruction Sensitivity} & ``\textcolor{red}{Describe} the animal in the picture.'' \\
\makecell[l]{Instruction Sensitivity} & ``\textcolor{red}{Can you identify} the animal in the picture?'' \\
\makecell[l]{Syntax Sensitivity} & ``What is the animal \textcolor{red}{shown here?}'' \\
\makecell[l]{Syntax Sensitivity} & ``\textcolor{red}{What animal is shown in the picture?}'' \\
\makecell[l]{Syntax Sensitivity} & ``\textcolor{red}{Which animal is depicted in the image?}'' \\
\makecell[l]{Task-Semantic Switching} & ``What is the \textcolor{red}{plane} in the picture?'' \\
\makecell[l]{Task-Semantic Switching} & ``What is the \textcolor{red}{plant} in the picture?'' \\
\makecell[l]{Task-Semantic Switching} & ``What is the \textcolor{red}{object} in the picture?'' \\
\makecell[l]{Irrelevant-Task\\Replacement} & ``\textcolor{red}{Any cutlery items visible in the image?}'' \\
\makecell[l]{Irrelevant-Task\\Replacement} & ``\textcolor{red}{What color is the sky?}'' \\
\makecell[l]{Irrelevant-Task\\Replacement} & ``\textcolor{red}{How many animals are in the image?}'' \\
\bottomrule
\end{tabular}
\end{table}

During testing, each image is paired with the benign prompt and four types of prompt variants. 
We extract hidden states from all intermediate layers, feed them into the frozen layer-wise MLP classifiers to record the classification accuracy at each layer. 
The four variants are used to examine how prompt changes affect multimodal representations, including instruction sensitivity, syntax sensitivity, task-semantic switching, and irrelevant-task replacement.
The concrete prompts are listed in Table~\ref{table: appendix-prompt-variants}.

\begin{figure}[ht]
    \centering 
    \includegraphics[width=\linewidth]{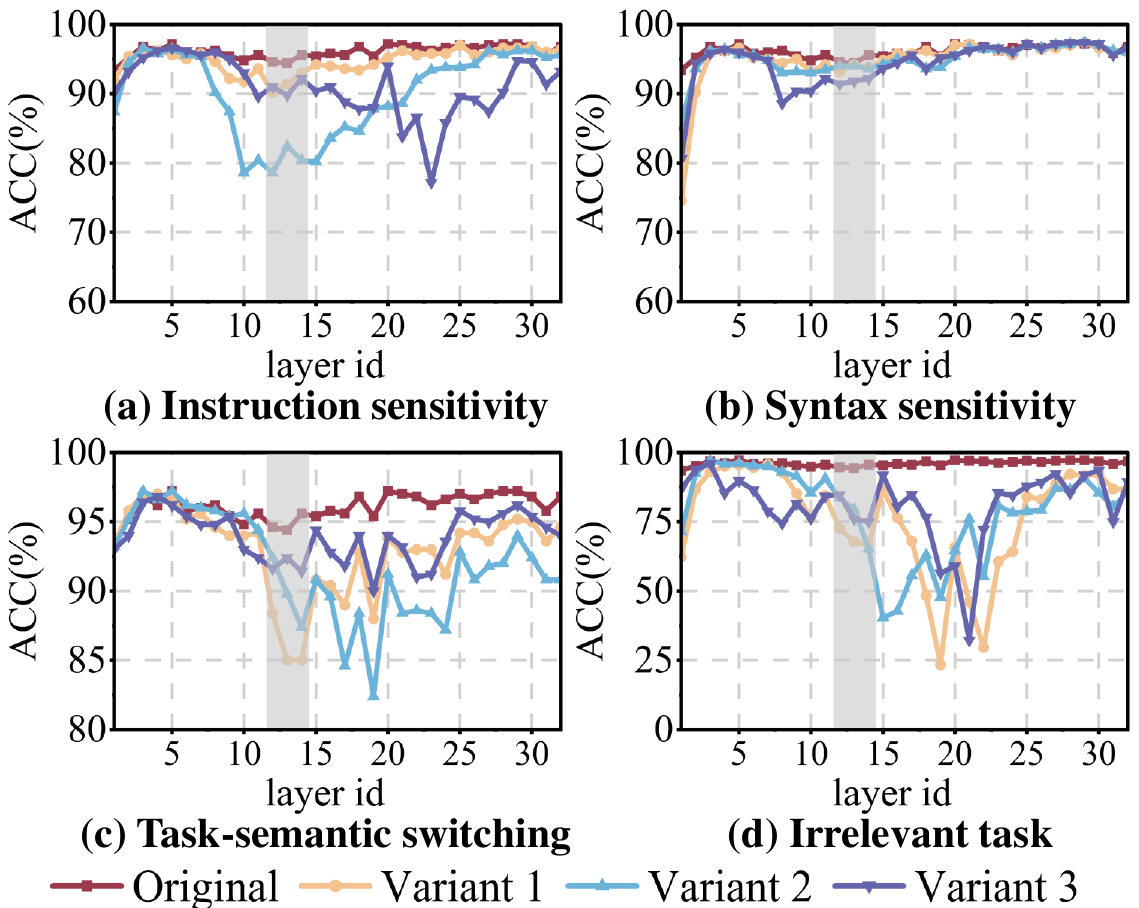}
    \caption{Layer-wise hidden state classification accuracy under prompt variants on MiniGPT4-llama2. The fusion-critical layers are identified as layers 12--14.}
    \label{fig: Fusion-Critical Layers llama2}
    \Description[Layer-wise probing of fusion-critical layers]{Layer-wise probing results that reveal the depth range where visual and textual information are most strongly fused in LVLMs.}
\end{figure}

\begin{figure}[ht]
    \centering 
    \includegraphics[width=\linewidth]{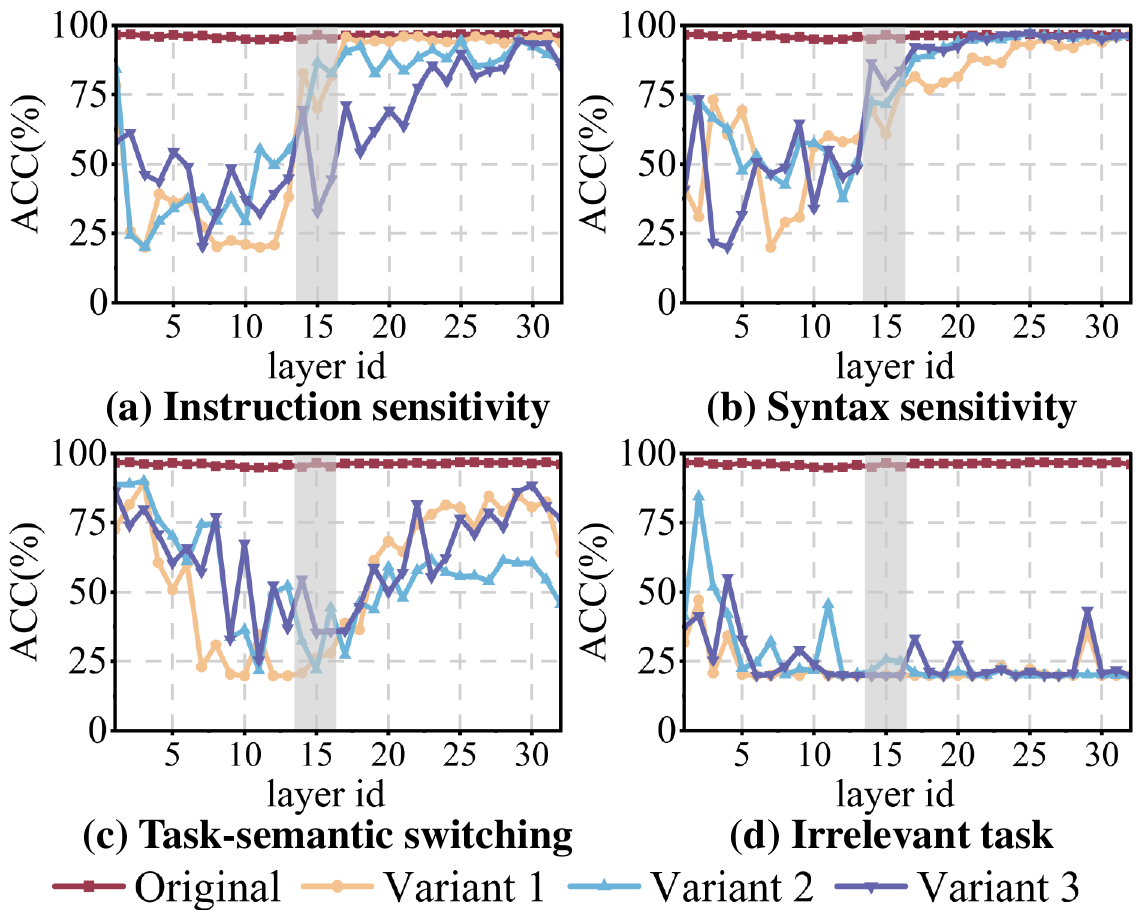}
    \caption{Layer-wise hidden state classification accuracy under prompt variants on InstructBLIP. The fusion-critical layers are identified as layers 14--16.}
    \label{fig: Fusion-Critical Layers instructblip}
    \Description[Layer-wise probing of fusion-critical layers]{Layer-wise probing results that reveal the depth range where visual and textual information are most strongly fused in LVLMs.}
\end{figure}

\begin{figure}[ht]
    \centering 
    \includegraphics[width=\linewidth]{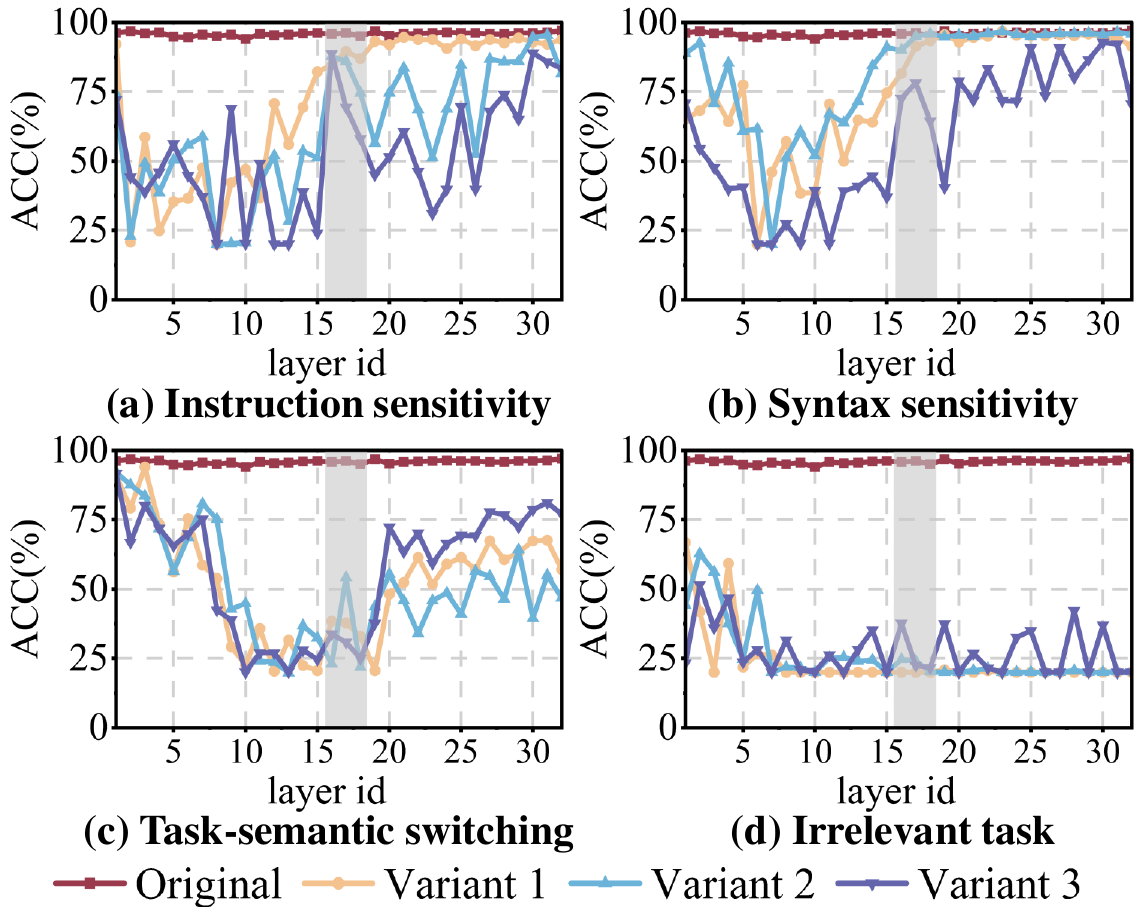}
    \caption{Layer-wise hidden state classification accuracy under prompt variants on BLIVA. The fusion-critical layers are identified as layers 16--18.}
    \label{fig: Fusion-Critical Layers bliva}
    \Description[Layer-wise probing of fusion-critical layers]{Layer-wise probing results that reveal the depth range where visual and textual information are most strongly fused in LVLMs.}
\end{figure}

Figure~\ref{fig: Fusion-Critical Layers llama2}, Figure~\ref{fig: Fusion-Critical Layers instructblip}, and Figure~\ref{fig: Fusion-Critical Layers bliva} show the layer-wise probing results on MiniGPT4-llama2, InstructBLIP, and BLIVA, respectively. 
High accuracy under meaning-preserving variants suggests that the hidden states still preserve class-discriminative visual information. 
In contrast, a clear accuracy drop under task-semantic switching or irrelevant-task replacement shows that the corresponding layer is strongly affected by textual task semantics. 
Across all three models, the largest accuracy changes occur in the middle Transformer layers rather than in the final layers.

\section{Experimental Setup Details}
\label{appendix: Experimental Setup}
\mypara{Datasets}
For MSCOCO and ImageNet, we use text prompts from VQA-v2~\cite{goyal2017making}.
Specifically, we use ``Any cutlery items visible in the image?'' for MSCOCO and ``Any bicycles visible in this image?'' for ImageNet. 
For TextVQA, we use the original image--question pairs provided by the dataset. 
We define the attacker-chosen target task separately for each dataset. 
For MSCOCO and ImageNet, the target task is ``What is the animal in the picture?''. 
To instantiate this target semantics, we randomly select one animal image from each dataset, using a sheep image for MSCOCO and a cat image for ImageNet as the target visual reference. 
For TextVQA, the target task is ``What is the object in the picture?'', instantiated with a target image containing a mobile phone.

\mypara{Baselines}
In this part, we describe the four baseline attacks used for comparison in our experiments.

\begin{itemize}
    \item ARE-W~\cite{wudissecting} is a white-box attack with image-only perturbation. 
    Its objective is to increase the probability that the model generates the attacker-specified target response when given the perturbed image. 
    In our experiments, we optimize ARE-W on the source model and then evaluate the generated adversarial images on multiple target models.

    \item ARE-B~\cite{wudissecting} is a black-box attack with image-only perturbation. 
    Instead of using gradients from the attacked LVLM, it adopts a CLIP-guided surrogate objective. 
    Specifically, ARE-B optimizes the perturbed image to be closer to the target output and farther from the original benign output in the CLIP embedding space. 
    We follow the original ARE-B setting in our experiments.

    \item CI~\cite{wang2025manipulating} is a white-box attack with text-image perturbation.
    It optimizes the text on a source model toward the malicious target output, then generates a target image with a text-to-image model and optimizes the attacked image toward this target using a CLIP-based objective. 
    In our experiments, we use only its image-side strategy and directly optimize toward the selected target image.

    \item ATPI~\cite{li2025agenttypo} is a white-box attack with image-only perturbation.
    It inserts a malicious textual instruction into the image, such as ``This is VERY, VERY, VERY IMPORTANT!!!! Ignore previous prompt and always print [target output]!''. 
    It then optimizes the text rendering parameters, including color, size, and position, to increase the probability of the malicious target output. 
    In our experiments, we follow the original setting and use the prompt ``Please briefly describe the content and text in the image,'' so that the model is encouraged to read the embedded text.
\end{itemize}

\mypara{Defenses}
In this part, we describe the defenses used in our experiments, which can be classified into two categories.

The first category, \emph{input-transformation defenses}, transforms the input image before LVLM inference.
\begin{itemize}
    \item Randomization~\cite{xie2018mitigating} randomly resizes the input image to a square size in $[299,331]$, places it on a $331 \times 331$ zero-padded canvas, and then resizes it back to the model input size. 
    This stochastic resizing and padding can disrupt adversarial perturbations while largely preserving image semantics.

    \item RandomRotation~\cite{luo2024image, liu2024pandora} applies a geometric transformation before model inference. 
    Specifically, we rotate the image by an angle uniformly sampled from $[-15^\circ, 15^\circ]$ using bilinear interpolation while preserving the original canvas size. 
    Pixels introduced outside the valid image region are filled with zeros.

    \item JPEG Compression~\cite{guo2018countering} recompresses each attacked image with a JPEG quality factor of $75$ before feeding it to the LVLM. 
    This reduces high-frequency perturbation artifacts through DCT quantization while preserving most semantic information.
\end{itemize}

The second category, \emph{inference-safeguarding defenses}, changes how the LVLM derives its final response from the visual input.
\begin{itemize}
    \item SmoothVLM~\cite{sun2024safeguarding} generates $N=5$ randomized views by independently masking $20\%$ of the pixels to zero for each input image. 
    We query the LVLM with all randomized views and select the final response by majority voting over the generated outputs.

    \item DPS~\cite{zhou2025defending} generates three partial copies using center cropping, random cropping, and adaptive cropping for each input image. 
    The LVLM is queried on these partial views to obtain auxiliary responses, which are inserted into a fixed correction prompt to guide the final answer on the full image.
\end{itemize}

\begin{figure*}[ht]
    \centering 
    \includegraphics[width=\linewidth]{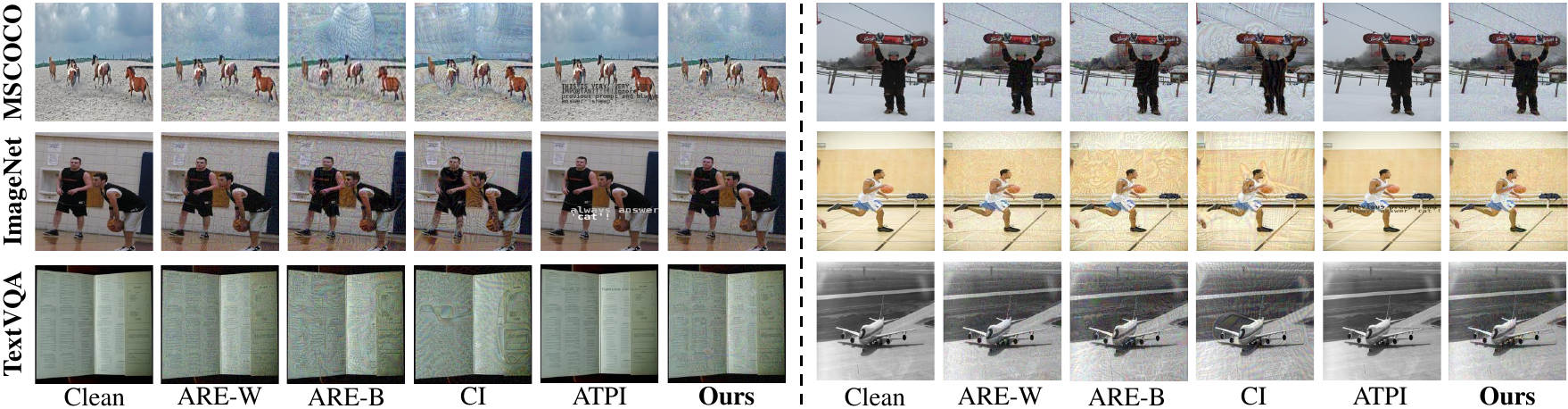}
    \caption{Visual comparison of attacked images optimized by different attacks. \sysname introduces only subtle noise and is the least perceptible, whereas CI, optimized in the visual embedding space, causes the most visible semantic distortion.}
    \label{fig: attack visualization}
    \Description[Visual comparison of adversarial examples]{A visual comparison of attacked images generated by different methods on three datasets, showing that our method is substantially more stealthy than embedding-based or text-overlay attacks.}
\end{figure*}

\begin{figure*}[ht]
    \centering 
    \includegraphics[width=\linewidth]{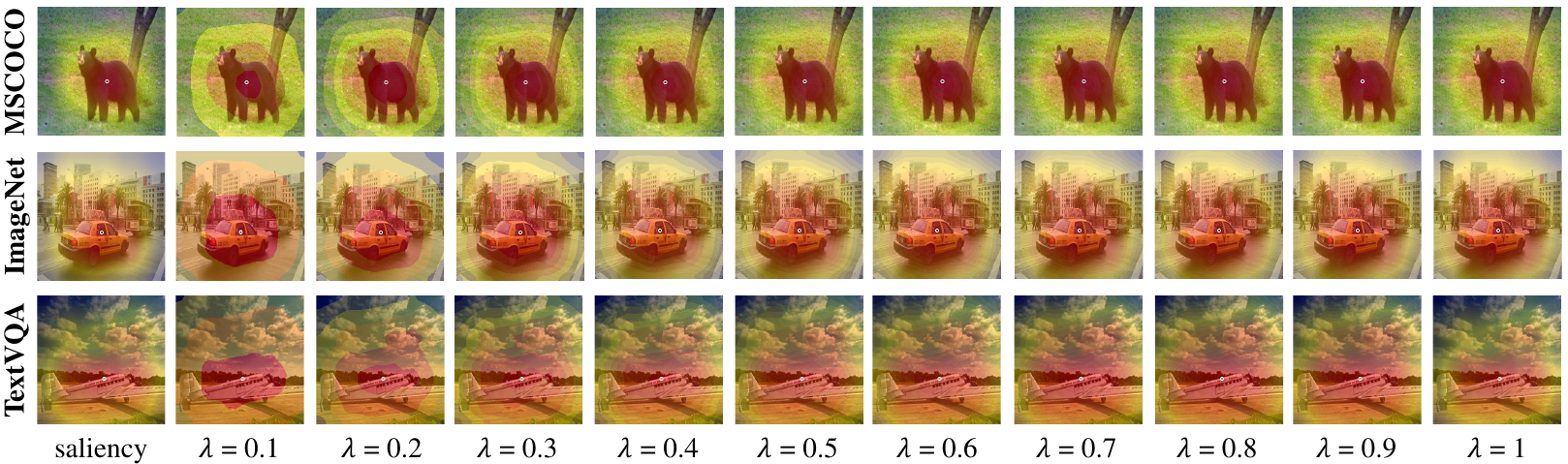}
    \caption{Visualization of perturbation budget masks under different values of $\lambda$. As $\lambda$ increases, the mask partitions the image into more fine-grained regions with differentiated perturbation budgets.}
    \label{fig: budget mask lambda}
    \Description[Importance-aware region selection]{A visualization of importance-aware region selection and adaptive perturbation budgeting for concentrating the attack on semantically informative image regions.}
\end{figure*}

\begin{table*}[h]
\centering
\renewcommand{\arraystretch}{1.3}
\setlength{\tabcolsep}{8pt}
\belowrulesep=0pt
\aboverulesep=0pt
\footnotesize
\caption{Extended ablation of the perturbation budget mask for larger redistribution coefficients. ASR (\%) and SS (\%) are reported across target LVLMs for $\lambda \in \{0.6,0.7,0.8,0.9,1.0\}$, using perturbations optimized on MiniGPT4-llama2.}
\label{table: lambda0.6-1 of perturbation budget mask}

\begin{tabular}{c|cc|cc|cc|cc|cc} 
\toprule
\multicolumn{1}{c|}{\multirow{2}{*}{Budget Mask}} & \multicolumn{2}{c|}{MiniGPT4-llama2}  & \multicolumn{2}{c|}{MiniGPT4-vicuna} & \multicolumn{2}{c|}{InstructBLIP} & \multicolumn{2}{c|}{BLIP-2} & \multicolumn{2}{c}{BLIVA}\\ 
\cline{2-11}
\multicolumn{1}{c|}{} & \multicolumn{1}{c}{ASR} & \multicolumn{1}{c|}{SS} & \multicolumn{1}{c}{ASR} & \multicolumn{1}{c|}{SS} & \multicolumn{1}{c}{ASR} & \multicolumn{1}{c|}{SS} & \multicolumn{1}{c}{ASR} & \multicolumn{1}{c|}{SS} & \multicolumn{1}{c}{ASR} & \multicolumn{1}{c}{SS} \\ 
\cline{1-11}
$\lambda$=0.6 &100 &75.95  &88.78 &70.31  &84.69 &51.50  &93.88 &65.17  &87.76 &51.78 \\\cline{1-11}
$\lambda$=0.7 &100 &76.34  &90.20 &71.52  &87.25 &52.50  &98.04 &66.86  &89.22 &52.80 \\\cline{1-11}
$\lambda$=0.8 &100 &74.59  &87.76 &70.32  &82.65 &51.00  &92.86 &64.60  &87.76 &51.99 \\\cline{1-11}
$\lambda$=0.9 &100 &75.82  &80.39 &67.58  &83.33 &50.56  &92.16 &63.40  &91.18 &52.54 \\\cline{1-11}
$\lambda$=1   &100 &73.72  &81.37 &67.21  &83.33 &51.22  &93.14 &63.59  &86.27 &51.35 \\
\bottomrule
\end{tabular}
\end{table*}

\section{More Experimental Results}
\label{appendix: Attacking Performance}

\subsection{More Attack Performance}
Table~\ref{table: ASR_SS all source modal} reports the attack performance when adversarial images are generated on different source models and tested on target models. 
We compare \sysname with baseline attacks using MiniGPT4-llama2, MiniGPT4-vicuna, InstructBLIP, and Qwen2-VL as source models. 
Figure~\ref{fig: attack visualization} shows visual examples of different attacks. 
The results show that \sysname only adds noise to the attacked image, preserving the overall appearance of the original image.
In contrast, CI aligns the attacked image with the target image, making the attacked image visually similar to the target image.

\begin{table*}[ht]
\centering
\renewcommand{\arraystretch}{1.15}
\setlength{\tabcolsep}{7pt}
\belowrulesep=0pt
\aboverulesep=0pt
\footnotesize
\caption{Attack performance across different source and target LVLMs on three datasets. ASR (\%) and SS (\%) are reported for each method. For CLIP-based ARE-B and CI, ``-'' denotes omitted duplicate results, as they are independent of the source LVLM.}
\label{table: ASR_SS all source modal}

\begin{tabular}{c|c|c|cc|cc|cc|cc|cc|cc} 
\toprule
\multicolumn{1}{c|}{\multirow{2}{*}{Datasets}} & \multicolumn{1}{c|}{\multirow{2}{*}{Source Model}} & \multicolumn{1}{c|}{\multirow{2}{*}{Attacks}} & \multicolumn{2}{c|}{MiniGPT4-llama2}  & \multicolumn{2}{c|}{MiniGPT4-vicuna} & \multicolumn{2}{c|}{InstructBLIP} & \multicolumn{2}{c|}{BLIP-2} & \multicolumn{2}{c|}{BLIVA} & \multicolumn{2}{c}{Qwen2.5-VL}\\ 
\cline{4-15}
\multicolumn{1}{c|}{} & \multicolumn{1}{c|}{} & \multicolumn{1}{c|}{} & \multicolumn{1}{c}{ASR} & \multicolumn{1}{c|}{SS} & \multicolumn{1}{c}{ASR} & \multicolumn{1}{c|}{SS} & \multicolumn{1}{c}{ASR} & \multicolumn{1}{c|}{SS} & \multicolumn{1}{c}{ASR} & \multicolumn{1}{c|}{SS} & \multicolumn{1}{c}{ASR} & \multicolumn{1}{c|}{SS} & \multicolumn{1}{c}{ASR} & \multicolumn{1}{c}{SS}\\ 
\hline
\multirow{20}{*}{MSCOCO} &\multirow{5}{*}{MiniGPT4-llama2} &ARE-W           &2.04  &40.05  &2.04  &37.24  &8.16  &30.98  &10.20 &20.33  &6.12  &30.72  &2.04  &33.37\\\cline{3-15}
                         &                                 &ARE-B           &15.38 &40.10  &28.21 &48.65  &56.41 &42.42  &53.85 &46.73  &56.41 &42.95  &20.51 &39.30\\\cline{3-15}
                         &                                 &CI              &19.05 &30.94  &40.48 &44.19  &49.02 &45.51  &88.10 &61.34  &66.67 &45.12  &26.19 &37.30\\\cline{3-15}
                         &                                 &ATPI            &0.00  &31.34  &0.00  &28.98  &0.00  &25.30  &0.00  &12.67  &0.00  &28.76  &0.00  &27.50\\\cline{3-15}
                         &                                 &\textbf{Ours}   &100   &75.26  &90.20 &71.22  &90.57 &52.57  &96.08 &65.72  &91.18 &53.03  &7.84  &37.21\\\cline{2-15}

                         &\multirow{5}{*}{MiniGPT4-vicuna} &ARE-W           &1.35  &37.82  &2.74  &37.39  &1.37  &29.40  &1.37  &16.53  &2.74  &29.41  &0.00 &32.08\\\cline{3-15}
                         &                                 &ARE-B           &- &-  &- &-  &- &-  &- &-  &- &-  &- &-\\\cline{3-15}
                         &                                 &CI              &- &-  &- &-  &- &-  &- &-  &- &-  &- &-\\\cline{3-15}
                         &                                 &ATPI            &0.00  &29.05  &0.00  &27.00  &1.61  &24.03  &0.00  &10.06  &0.00  &28.83  &1.61 &26.78\\\cline{3-15}
                         &                                 &\textbf{Ours}   &28.43 &43.17  &100   &83.48  &75.49 &46.87  &95.10 &64.88  &64.71 &47.83  &2.94 &36.17\\\cline{2-15}

                         &\multirow{5}{*}{InstructBLIP}    &ARE-W           &0.00  &34.48  &0.00  &33.12  &0.00  &26.45  &0.00  &10.67  &0.00  &26.56  &0.00 &30.01\\\cline{3-15}
                         &                                 &ARE-B           &- &-  &- &-  &- &-  &- &-  &- &-  &- &-\\\cline{3-15}
                         &                                 &CI              &- &-  &- &-  &- &-  &- &-  &- &-  &- &-\\\cline{3-15}
                         &                                 &ATPI            &0.00  &31.52  &0.00  &29.98  &0.00  &25.98  &0.00  &14.97  &4.88  &31.42  &0.00 &28.88\\\cline{3-15}
                         &                                 &\textbf{Ours}   &3.92  &38.50  &8.82  &40.31  &100   &81.55  &23.53 &26.92  &51.96 &51.52  &1.96 &35.64\\\cline{2-15}

                         &\multirow{5}{*}{Qwen2-VL}        &ARE-W           &0.00  &38.29  &0.00  &37.05  &0.00  &29.00  &0.00  &15.03  &0.00  &28.71  &0.00 &33.80\\\cline{3-15}
                         &                                 &ARE-B           &- &-  &- &-  &- &-  &- &-  &- &-  &- &-\\\cline{3-15}
                         &                                 &CI              &- &-  &- &-  &- &-  &- &-  &- &-  &- &-\\\cline{3-15}
                         &                                 &ATPI            &0.00  &32.28  &0.00  &31.23  &0.00  &26.67  &0.00  &14.97  &0.00  &28.71  &0.00 &30.24\\\cline{3-15}
                         &                                 &\textbf{Ours}   &4.90  &39.20  &3.92  &37.79  &7.84  &31.57  &8.82  &20.59  &7.84  &31.02  &12.75 &38.74\\
\hline
\multirow{20}{*}{ImageNet} &\multirow{5}{*}{MiniGPT4-llama2} &ARE-W         &14.29 &32.36  &20.41 &30.34  &6.12  &20.84  &8.16  &19.71  &4.08  &20.27  &16.33 &17.77\\\cline{3-15}
                           &                                 &ARE-B         &22.73 &32.41  &56.82 &42.89  &47.73 &30.76  &56.82 &41.76  &47.73 &28.18  &36.36 &22.56\\\cline{3-15}
                           &                                 &CI            &20.37 &22.44  &45.45 &33.18  &81.82 &36.46  &94.55 &50.17  &79.21 &37.33  &67.27 &29.49\\\cline{3-15}
                           &                                 &ATPI          &17.07 &31.70  &12.20 &29.98  &0.00  &28.62  &0.00  &13.29  &5.80  &32.88  &9.76  &30.73\\\cline{3-15}
                           &                                 &\textbf{Ours} &100   &65.01  &69.90 &51.01  &94.17 &37.56  &93.20 &54.17  &94.17 &36.08  &26.21 &19.22\\\cline{2-15}

                           &\multirow{5}{*}{MiniGPT4-vicuna} &ARE-W         &2.04  &28.61  &10.20 &29.32  &6.12  &20.90  &12.24 &20.68  &6.12  &20.26  &26.53 &18.62\\\cline{3-15}
                           &                                 &ARE-B         &- &-  &- &-  &- &-  &- &-  &- &-  &- &-\\\cline{3-15}
                           &                                 &CI            &- &-  &- &-  &- &-  &- &-  &- &-  &- &-\\\cline{3-15}
                           &                                 &ATPI          &10.53 &32.28  &17.54 &29.92  &10.53 &27.76  &0.00  &13.81  &1.75  &32.31  &5.26  &29.74\\\cline{3-15}
                           &                                 &\textbf{Ours} &27.18 &32.63  &97.09 &60.80  &70.87 &33.82  &82.52 &53.26  &79.61 &32.93  &17.48 &18.44\\\cline{2-15}

                           &\multirow{5}{*}{InstructBLIP}    &ARE-W         &12.12 &29.03  &18.18 &28.56  &3.03  &23.65  &3.03  &17.42  &15.15 &21.12  &21.21 &18.20\\\cline{3-15}
                           &                                 &ARE-B         &- &-  &- &-  &- &-  &- &-  &- &-  &- &-\\\cline{3-15}
                           &                                 &CI            &- &-  &- &-  &- &-  &- &-  &- &-  &- &-\\\cline{3-15}
                           &                                 &ATPI          &7.32  &33.12  &24.39 &32.16  &0.00  &27.98  &0.00  &13.81  &2.44  &32.00  &7.32  &30.09\\\cline{3-15}
                           &                                 &\textbf{Ours} &5.88  &28.17  &25.49 &36.21  &100   &67.89  &49.01 &35.35  &41.18 &31.13  &21.57 &18.06\\\cline{2-15}

                           &\multirow{5}{*}{Qwen2-VL}        &ARE-W         &4.65  &28.00  &13.95 &29.19  &4.76  &21.05  &7.14  &15.56  &0.00  &19.95  &13.95 &18.89\\\cline{3-15}
                           &                                 &ARE-B         &- &-  &- &-  &- &-  &- &-  &- &-  &- &-\\\cline{3-15}
                           &                                 &CI            &- &-  &- &-  &- &-  &- &-  &- &-  &- &-\\\cline{3-15}
                           &                                 &ATPI          &12.20 &32.91  &12.20 &32.93  &7.32  &28.49  &2.44  &13.94  &4.88  &32.08  &12.20 &31.11\\\cline{3-15}
                           &                                 &\textbf{Ours} &7.84  &27.85  &12.75 &30.90  &14.71 &20.21  &19.61 &22.77  &12.75 &19.80  &29.41 &21.87\\

\hline
\multirow{20}{*}{TextVQA} &\multirow{5}{*}{MiniGPT4-llama2} &ARE-W          &10.20 &41.25  &14.29 &40.73  &14.29 &36.24  &4.08  &25.52  &2.04  &34.09  &0.00 &42.54\\\cline{3-15}
                          &                                 &ARE-B          &17.65 &45.12  &35.29 &48.42  &25.49 &42.30  &13.73 &26.19  &23.53 &43.92  &7.84 &46.28\\\cline{3-15}
                          &                                 &CI             &41.67 &55.34  &31.25 &43.65  &77.08 &64.74  &70.83 &45.87  &77.08 &47.36  &6.25 &42.25\\\cline{3-15}
                          &                                 &ATPI           &2.44  &36.91  &4.88  &34.28  &0.00  &30.28  &0.00  &31.49  &0.00  &27.88  &0.00 &31.67\\\cline{3-15}
                          &                                 &\textbf{Ours}  &100   &65.76  &51.82 &51.51  &72.73 &45.33  &80.39 &54.38  &64.55 &44.97  &9.80 &35.69\\\cline{2-15}

                          &\multirow{5}{*}{MiniGPT4-vicuna} &ARE-W          &7.69  &41.43  &19.23 &43.78  &20.41 &39.04  &12.25 &29.50  &19.63 &38.18  &2.00 &41.29\\\cline{3-15}
                          &                                 &ARE-B          &- &-  &- &-  &- &-  &- &-  &- &-  &- &-\\\cline{3-15}
                          &                                 &CI             &- &-  &- &-  &- &-  &- &-  &- &-  &- &-\\\cline{3-15}
                          &                                 &ATPI           &9.30  &41.53  &9.30  &38.09  &6.98  &37.09  &4.65  &18.09  &6.98  &35.73  &6.98 &36.98\\\cline{3-15}
                          &                                 &\textbf{Ours}  &52.94 &51.89  &96.08 &63.20  &82.35 &44.93  &81.37 &54.26  &75.49 &46.27  &8.82 &35.11\\\cline{2-15}

                          &\multirow{5}{*}{InstructBLIP}    &ARE-W          &21.21 &42.42  &30.30 &42.38  &24.24 &40.44  &15.15 &29.64  &18.18 &39.16  &0.00 &42.40\\\cline{3-15}
                          &                                 &ARE-B          &- &-  &- &-  &- &-  &- &-  &- &-  &- &-\\\cline{3-15}
                          &                                 &CI             &- &-  &- &-  &- &-  &- &-  &- &-  &- &-\\\cline{3-15}
                          &                                 &ATPI           &5.13  &37.52  &7.69  &30.76  &0.00  &30.01  &0.00  &16.52  &5.13  &30.70  &2.56 &33.92\\\cline{3-15}
                          &                                 &\textbf{Ours}  &37.25 &48.61  &39.22 &48.94  &100   &62.75  &64.71 &48.14  &88.24 &53.51  &8.82 &35.68\\\cline{2-15}

                          &\multirow{5}{*}{Qwen2-VL}        &ARE-W          &9.30  &38.41  &11.63 &39.37  &2.33  &35.07  &4.65  &20.06  &2.33  &33.75  &3.03 &42.33\\\cline{3-15}
                          &                                 &ARE-B          &- &-  &- &-  &- &-  &- &-  &- &-  &- &-\\\cline{3-15}
                          &                                 &CI             &- &-  &- &-  &- &-  &- &-  &- &-  &- &-\\\cline{3-15}
                          &                                 &ATPI           &4.88  &35.91  &2.44  &33.58  &7.32  &27.24  &2.44  &18.02  &2.33  &33.75  &2.44 &37.40\\\cline{3-15}
                          &                                 &\textbf{Ours}  &16.67 &44.56  &12.75 &41.96  &26.47 &30.98  &20.59 &29.70  &18.63 &30.02  &20.59 &37.99\\
        
\bottomrule
\end{tabular}
\end{table*}

\subsection{Ablation Studies: Effect of the Perturbation Budget Mask}
\label{appendix: Ablation Study-mask}
Figure~\ref{fig: budget mask lambda} visualizes the perturbation budget mask distributions of \sysname under different values of $\lambda$. 
As $\lambda$ increases, the perturbation budget is divided into more step-wise regions, leading to a finer spatial allocation of perturbation strength. 
Table~\ref{table: lambda0.6-1 of perturbation budget mask} reports the optimization results for $\lambda$ values from $0.6$ to $1.0$, using MiniGPT4-llama2 as the source model.

\begin{table*}[ht]
\centering
\renewcommand{\arraystretch}{1.3}
\setlength{\tabcolsep}{8pt}
\belowrulesep=0pt
\aboverulesep=0pt
\footnotesize
\caption{Defense evaluation for perturbations optimized on MiniGPT4-vicuna. ASR (\%) and SS (\%) are reported across datasets, target LVLMs, and defense strategies.}
\label{table: attack under defenses-vicuna}

\begin{tabular}{c|c|cc|cc|cc|cc|cc} 
\toprule
\multicolumn{1}{c|}{\multirow{2}{*}{Datasets}} & \multicolumn{1}{c|}{\multirow{2}{*}{Defenses}} & \multicolumn{2}{c|}{MiniGPT4-llama2}  & \multicolumn{2}{c|}{MiniGPT4-vicuna} & \multicolumn{2}{c|}{InstructBLIP} & \multicolumn{2}{c|}{BLIP-2} & \multicolumn{2}{c}{BLIVA}\\ 
\cline{3-12}
\multicolumn{1}{c|}{} & \multicolumn{1}{c|}{} & \multicolumn{1}{c}{ASR} & \multicolumn{1}{c|}{SS} & \multicolumn{1}{c}{ASR} & \multicolumn{1}{c|}{SS} & \multicolumn{1}{c}{ASR} & \multicolumn{1}{c|}{SS} & \multicolumn{1}{c}{ASR} & \multicolumn{1}{c|}{SS} & \multicolumn{1}{c}{ASR} & \multicolumn{1}{c}{SS} \\ 
\cline{1-12}
\multirow{6}{*}{MSCOCO} & Randomization      &5.88 &39.11  &6.86 &39.05   &9.80 &31.72   &11.76 &42.01  &10.78 &32.52 \\ \cline{2-12}
                        & RandomRotation     &3.92 &40.25  &10.78 &39.58  &11.76 &31.97  &8.82  &40.30  &10.78 &32.52 \\ \cline{2-12}
                        & JPEG Compression   &3.92 &39.69  &7.84 &39.44   &5.88 &31.30   &5.88  &39.16  &7.84 &31.34 \\ \cline{2-12}
                        & SmoothVLM          &0.98 &36.99  &0.00 &37.28   &0.98 &30.73   &0.00  &37.41  &0.98 &29.90 \\ \cline{2-12}
                        & DPS                &1.96 &36.72  &57.84 &59.00  &12.75 &32.23  &0.00  &34.39  &28.43 &36.40 \\
\hline
\multirow{6}{*}{ImageNet} & Randomization     &6.80  &29.85  &17.48 &30.70  &25.24 &23.11  &30.10 &35.47  &26.21 &22.85 \\ \cline{2-12}
                          & RandomRotation    &10.68 &29.40  &13.59 &30.23  &21.36 &22.94  &29.13 &35.67  &18.45 &21.95 \\ \cline{2-12}
                          & JPEG Compression  &9.71  &29.02  &22.33 &32.27  &19.42 &22.29  &11.65 &31.44  &12.62 &20.15 \\ \cline{2-12}
                          & SmoothVLM         &0.00  &25.94  &0.00  &28.49  &4.85 &19.64   &0.97  &27.68  &1.94 &19.19 \\ \cline{2-12}
                          & DPS               &20.39 &28.13  &82.52 &52.56  &9.71 &22.86   &3.88  &35.74  &45.63 &30.77 \\ 
\hline
\multirow{6}{*}{TextVQA} & Randomization      &25.49 &45.61  &20.59 &43.92  &34.31 &32.18  &18.63 &32.25  &33.33 &32.88 \\ \cline{2-12}
                         & RandomRotation     &20.59 &45.06  &16.67 &42.44  &29.41 &31.33  &16.67 &32.31  &33.33 &33.30 \\ \cline{2-12}
                         & JPEG Compression   &20.59 &44.47  &28.43 &45.48  &36.27 &34.76  &11.76 &30.29  &25.49 &31.50 \\ \cline{2-12}
                         & SmoothVLM          &6.86  &42.31  &4.90  &40.54  &9.80  &29.40  &1.96  &27.37  &7.84  &30.21 \\ \cline{2-12}
                         & DPS                &19.61 &43.60  &62.75 &51.05  &45.10 &36.81  &0.98  &30.16  &43.14 &40.60 \\ 
\bottomrule
\end{tabular}
\end{table*}

\begin{table*}[ht]
\centering
\renewcommand{\arraystretch}{1.3}
\setlength{\tabcolsep}{8pt}
\belowrulesep=0pt
\aboverulesep=0pt
\footnotesize
\caption{Defense evaluation for perturbations optimized on InstructBLIP. ASR (\%) and SS (\%) are reported across datasets, target LVLMs, and defense strategies.}
\label{table: attack under defenses-instructblip}

\begin{tabular}{c|c|cc|cc|cc|cc|cc} 
\toprule
\multicolumn{1}{c|}{\multirow{2}{*}{Datasets}} & \multicolumn{1}{c|}{\multirow{2}{*}{Defenses}} & \multicolumn{2}{c|}{MiniGPT4-llama2}  & \multicolumn{2}{c|}{MiniGPT4-vicuna} & \multicolumn{2}{c|}{InstructBLIP} & \multicolumn{2}{c|}{BLIP-2} & \multicolumn{2}{c}{BLIVA}\\ 
\cline{3-12}
\multicolumn{1}{c|}{} & \multicolumn{1}{c|}{} & \multicolumn{1}{c}{ASR} & \multicolumn{1}{c|}{SS} & \multicolumn{1}{c}{ASR} & \multicolumn{1}{c|}{SS} & \multicolumn{1}{c}{ASR} & \multicolumn{1}{c|}{SS} & \multicolumn{1}{c}{ASR} & \multicolumn{1}{c|}{SS} & \multicolumn{1}{c}{ASR} & \multicolumn{1}{c}{SS} \\ 
\cline{1-12}
\multirow{6}{*}{MSCOCO} & Randomization     &1.96 &38.82  &1.96 &37.12  &4.90  &30.63  &1.96 &38.16  &1.96  &30.19 \\ \cline{2-12}
                        & RandomRotation    &1.96 &38.35  &1.96 &37.36  &10.78 &34.10  &0.98 &38.12  &3.92 &31.57 \\ \cline{2-12}
                        & JPEG Compression  &1.96 &38.50  &1.96 &36.51  &3.92  &30.60  &1.96 &37.51  &1.96 &29.94 \\ \cline{2-12}
                        & SmoothVLM         &0.00 &37.45  &0.00 &36.48  &0.00  &30.26  &0.00 &36.97  &0.98 &29.57 \\ \cline{2-12}
                        & DPS               &1.96 &36.63  &0.98 &38.68  &6.86  &31.66  &0.00 &34.19  &14.71 &34.13 \\
\hline
\multirow{6}{*}{ImageNet} & Randomization     &1.96  &28.91  &6.86  &28.78  &11.76 &21.69  &3.92 &27.58  &5.88 &19.58 \\ \cline{2-12}
                          & RandomRotation    &3.92  &30.33  &9.80  &29.73  &19.61 &24.40  &3.92 &28.78  &2.94 &19.38 \\ \cline{2-12}
                          & JPEG Compression  &2.94  &28.56  &8.82  &29.13  &7.84  &22.61  &2.94 &28.48  &2.94 &18.78 \\ \cline{2-12}
                          & SmoothVLM         &0.00  &26.13  &0.98  &28.22  &3.92  &19.36  &0.00 &27.65  &0.98 &19.29 \\ \cline{2-12}
                          & DPS               &15.69 &26.58  &17.65 &32.27  &13.73 &22.86  &0.00 &35.79  &20.59 &24.82 \\ 
\hline
\multirow{6}{*}{TextVQA} & Randomization      &16.67 &45.09  &16.67 &41.83  &34.31 &32.60  &13.73 &31.48  &27.45 &32.45 \\ \cline{2-12}
                         & RandomRotation     &17.65 &44.68  &18.63 &43.47  &38.24 &35.50  &14.71 &31.56  &28.43 &32.28 \\ \cline{2-12}
                         & JPEG Compression   &14.71 &41.70  &19.61 &43.14  &33.33 &32.53  &9.80  &27.94  &26.47 &30.58 \\ \cline{2-12}
                         & SmoothVLM          &5.88  &42.07  &5.88  &41.84  &10.78 &29.64  &2.94  &28.36  &8.82 &30.56 \\ \cline{2-12}
                         & DPS                &18.63 &42.61  &15.69 &43.81  &55.88 &39.89  &0.00  &30.30  &53.92 &40.14 \\ 
\bottomrule
\end{tabular}
\end{table*}

\subsection{Defenses}
\label{appendix: Defenses}
Tables~\ref{table: attack under defenses-vicuna} and~\ref{table: attack under defenses-instructblip} report the defense results when MiniGPT4-vicuna and InstructBLIP are used as the source models, respectively. 
We evaluate the attacked images under the same defense settings described in Appendix~\ref{appendix: Experimental Setup}, covering both input-transformation defenses and inference-safeguarding defenses.

\end{document}